\documentclass[11pt,preprint]{aastex}

\newcommand{\simgt}{\;\hbox{\rlap{\raise 0.425ex\hbox{$>$}}\lower 0.65ex\hbox{$\sim$}}\;}
\newcommand{\simlt}{\;\hbox{\rlap{\raise 0.425ex\hbox{$<$}}\lower 0.65ex\hbox{$\sim$}}\;}

\begin{document}

%\baselineskip9pt
%\tableofcontents
%\baselineskip13pt
%\newpage

% \title{Lower Limit on Dark Matter Collisionality from Abell 3827}
%\title{Galaxy/Lens-mass Offsets in Abell 3827:\\ Collisional Dark Matter?}
\title{Light/Mass Offsets in the Lensing Cluster Abell 3827: \\
Evidence for Collisional Dark Matter?}

\author{ 
Liliya L. R. Williams\altaffilmark{1},
Prasenjit Saha\altaffilmark{2}
}

\altaffiltext{1}{Department of Astronomy, University of Minnesota, 
116 Church Street SE, Minneapolis, MN 55455, USA; llrw@astro.umn.edu}
\altaffiltext{2}{Institute for Theoretical Physics, University of Z\"urich, 
Winterthurerstrasse 190, CH-8057 Z\"urich, Switzerland; psaha@physik.uzh.ch}

\begin{abstract}
If dark matter has a non-zero self-interaction cross-section, then
dark matter halos of individual galaxies in cluster cores should
experience a drag force from the ambient dark matter of the cluster,
which will not affect the stellar components of galaxies, and thus
will lead to a separation between the stellar and dark matter.  If the
cross-section is only a few decades below its current astrophysically
determined upper limit, then kpc-scale separations should
result. However, such separations will be observable only under very
favorable conditions. Abell 3827 is a nearby late stage cluster merger
with four massive central ellipticals within $20$~kpc of each other.
The ten strong lensing images tightly surrounding the ellipticals
provide an excellent set of constraints for a free-form lens
reconstruction. Our free-form mass maps show a massive dark extended
clump, about $6$~kpc from one of the ellipticals. The robustness of
this result has been tested with many reconstructions, and confirmed
with experiments using synthetic lens mass distributions. Interpreted
in terms of dark matter collisionality, our result yields
$\sigma/m\simgt 4.5\; 10^{-7}\,(t/10^{10}{\rm yr})^{-2}$~cm$^2$~g$^{-1}$,
where $t$ is the merger's age.
\end{abstract}

\section{Introduction}\label{intro}

Constraints on dark matter self-scattering cross-section can be
obtained from astrophysical observations.  Upper limits on the
cross-section per unit mass, $\sigma/m$ have been computed, most
famously, from the Bullet Cluster \citep{clowe06} which is a plane of
the sky collision between two galaxy clusters. The distribution of
mass was measured using lensing, and that of the gas using
observations of its X-ray emission. From the consistency of the
mass-to-light ratios of the sub-cluster and the main cluster, which
implies that very little dark matter was lost due to self-interaction
during the merger, \cite{mark04} derive an upper limit on the
cross-section, as $\sigma/m<1$ cm$^2$~g$^{-1}$. Based on the same
argument, but using $N\!$-body simulations to model the collision,
\cite{rand08} derive $\sigma/m<0.7$ cm$^2$~g$^{-1}$.  The authors note
that because the separation on the sky between the centroid of the
visible galaxies and dark matter is consistent with zero, the dark
matter's interaction cross-section is $\sigma/m<1.25$ cm$^2$~g$^{-1}$.

Relaxed clusters offer a different route to the upper limit on
$\sigma/m$. Random scattering between dark matter particles will tend
to make the dense cluster core more spherical. \cite{me02} used an
elliptical core of MS2137-23 to place an upper limit of $0.02$
cm$^2$~g$^{-1}$.  The value relies on the assumption that the measured 
ellipticity comes solely from  the cluster core, instead of projection 
of mass from outside the core, and that there has been no recent merger 
activity in the cluster. This is the most stringent upper limit on 
$\sigma/m$ from astrophysical observations.

Clusters also present a potential opportunity to measure a lower limit
on $\sigma/m$ (c.f.~\cite{deL95})  Consider galaxies in the central regions of a
cluster. These galaxies, usually ellipticals, consist of a compact
stellar component and a more extended dark matter halo. As galaxies
orbit the cluster centre, they move through the ambient dark matter
halo of the cluster.  The galaxies' stellar matter and dark matter
interact with the ambient dark matter differently.  The motion of
stars is subject to the gravitational field of the cluster and is
unaffected by individual dark matter particles. The galaxies' dark
matter particles, on the other hand, can interact and scatter off the
cluster's dark matter, thereby retarding the forward motion of
galaxy's dark matter, making it lag behind the stellar component of
the galaxy. If the scattering cross-section and the density of dark
matter are large enough, this 'drag' force will eventually result in a
spatial separation between the stellar and dark matter components of
the galaxy, which might be detectable.

%One can think of the dark matter within the galaxies as 'gas' with
%some self interaction cross-section per unit mass, $\sigma/m$. Ram
%pressure due to the dark matter of the main cluster or other nearby
%galaxies will act on the dark matter particles of the galaxy in
%question and retard its forward motion. The situation is akin to ram
%pressure effects on gas of a galaxy going through a cluster. In that
%case, one usually estimates the ram pressure,
%$P_{ram}=\rho_{gas}v^2_{gal}$, and the 'gravitational' pressure,
%$P_{grav}=\alpha\,{GM{gal}(r)\rho_{gas}(r)}/{r}$, with $\alpha$ a
%fudge factor of order unity. $P_{ram}=P_{grav}$ gives the stripping
%radius beyond which all gas is assumed to be stripped.

One can quantify the above scenario and estimate the separation as a
function of relevant parameters. Assume that the mass of the core of
the cluster interior to the centre of the galaxy is $M_{co}$, the dark
matter and the stellar masses of the galaxy are $M_{dm}$ and $M_{st}$,
respectively, and their average distance from the core are,
$r_{st}\approx r_{dm}\approx r$.  Let the cross-sectional area of the
galaxy's dark matter halo be $\pi s^2$.  The force of gravity between
the cluster core and the two components of the galaxy are,
respectively,
\begin{equation}
F_{st}\sim\frac{GM_{co}M_{st}}{r^2}\label{newton}
\end{equation}
\begin{equation}
F_{dm}\sim\frac{GM_{co}M_{dm}}{r^2}\times\Bigl[1-\frac{M_{dm}\,\sigma/m}{\pi
s^2}\Bigr].\label{newtDM}
\end{equation}
In the second equation gravity is partly counteracted, or screened by
the self-interaction and scattering between the core's and galaxy's
dark matter particles. The quantity $({M_{dm}\,\sigma/m})/({\pi s^2})$
is the covering factor, or the `optical depth', and is assumed to be
small.  Over time, the difference in the two accelerations amounts to
a displacement, or separation
\begin{equation}
d\sim
\Bigl(\frac{F_{st}}{M_{st}}-\frac{F_{dm}}{M_{dm}}\Bigr)\;t^2 =
\frac{GM_{co}M_{dm}\,\sigma/m}{\pi r^2 s^2}\;t^2.\label{disp}
\end{equation}
If (i) the galaxy orbit were exactly radial, (ii) the cluster's dark
matter space density scaled as $\rho(r)\sim r^{-1}$, so $M_{co}/r^2$
were independent of radius at all epochs, (iii) the spatial
distribution of stars and dark matter within the galaxy were
point-like, and (iv) the expression in the square brackets in
eq.~\ref{newtDM} were an appropriate way to quantify the retarding
force, then the displacement would be exactly $d/2$.  Of these,
assumption (iv) is the least objectionable and will lead
to a relatively small error.
Assumption (iii), if it is not valid and dark matter halos are
actually extended and overlapping, will tend to underestimate the
force in Equation~(\ref{newtDM}) and hence the separation $d$.  Then a
larger cross-section would be required to produce a given
displacement. Hence a point-like mass distributions is conservative,
if a lower limit on $\sigma/m$ is being sought.  Assumptions (i)
and (ii), and the $t^2$ dependence of $d$ in eq.~\ref{disp} combined
summarize the dynamical evolution of the cluster and the galaxy's
orbit, and are not necessarily independent of each other. For a
typical relaxed cluster the central density profile is not too far
from $\rho(r)\sim r^{-1}$, however, it is not expected to be exactly
that, and would have most likely varied over time. Consequently, the
shape of the galaxy's orbit and how it evolved in the past is
important, and would constitute the largest source of uncertainty in
$d$.

Equation~\ref{disp} can be rewritten as
\begin{equation}
d = 6\times 10^4
\Bigl(\frac{M_{co}}{10^{12}M_\odot}\Bigr)
\Bigl(\frac{M_{dm}}{10^{11}M_\odot}\Bigr)
\Bigl(\frac{\sigma/m}{0.02~{\rm{cm}^2~\rm{g}^{-1}}}\Bigr)
\Bigl(\frac{s}{10~\rm{kpc}}\Bigr)^{-2}
\Bigl(\frac{r}{10~\rm{kpc}}\Bigr)^{-2}
\Bigl(\frac{t}{10^{10}~\rm{yr}}\Bigr)^{2}
~\rm{kpc}
\label{dispunits}
\end{equation}
Thus, for typical values of masses and distances, and assuming that
the self-scattering cross-section is at most a few decades below the
most stringent astrophysically obtained upper limit, the separation
between the dark and visible components of galaxies will be or order
of kpc, and hence potentially observable.

Our order-of-magnitude estimate can be compared to the results of
$N\!$-body simulations carried out by \cite{rand08}, who modeled the
Bullet Cluster. For the specific geometry and parameters of that
cluster, and assuming $\sigma/m=0.02$ cm$^2$~gm$^{-1}$, the separation
between the stellar and dark matter in the sub-cluster can be read off
from their Figure~5, as $\simlt 2$~kpc. This is smaller than the
separation we get for the parameters assumed in eq.~\ref{dispunits}.
The main difference between the two estimates is due to the nature of
the encounters in the two cases. In the Bullet Cluster, the two
merging clusters made one passage past each other, so that the
sub-cluster spent little time in the dense core region of the larger
cluster. For the Bullet Cluster, the current separation of the
sub-components is 720 kpc. A merger velocity of 4700 km~s$^{-1}$
\citep[following][]{rand08} gives an encounter time of $2.3\;
10^{8}$~yrs.  The velocity may well be significantly lower
\citep{sf07}.  But \cite{rand08} report that velocity of $\sim$ 2800
km~s$^{-1}$ would not greatly change their results.

In the scenario we outlined above, the galaxies have been orbiting the
inner cluster core for about a Hubble time. Also, $r\sim 100$~kpc is a
more appropriate value for the Bullet Cluster. Taking these in to
account, eq.~\ref{dispunits} gives $d\sim 0.1$~kpc, comparable to
their estimate of $\simlt 2$~kpc. Thus, the Bullet Cluster itself is a
(limited) test of eq.~\ref{dispunits}.

This comparison with the Bullet Cluster demonstrates that the longer
duration of the encounter and smaller distances in our scenario give
rise to a larger separation, more likely to be detected.

\section{The Cluster Abell 3827}\label{clus}

Abell 3827 is rather unique; it is a nearby example of a multiple
galactic cannibalism in progress, where the massive central
ellipticals are closely surrounded by lensed images.  The cluster's
dynamical state, its proximity to us, and the fortuitous lensing
configuration make it a prime candidate for detecting any displacement
between the dark and visible matter due to non-zero dark matter
scattering cross-section.  % moved upstream

In this paper we use strongly lensed images to reconstruct the mass
distribution in the core of Abell 3827. As we will show, the core
contains a galaxy whose visible and total mass are apparently
separated by a few kpc. One possible interpretation of this result is
the scenario described in the previous Section. The cluster's lensing
features were recently discovered by \cite{car10}, whose findings we
summarize below.

Abell 3827 ($z\approx 0.1$) appears to be in the last stages of merger
\citep{car10}.  The outer cluster looks smooth and relaxed, and there
are no significant secondary galaxy concentrations. The core of the
cluster contains five ellipticals, called $N1-N5$, within $\sim
15$~kpc of the centre, which appear to be in the process of spiraling
inwards to form a single central galaxy. These ellipticals have
probably undergone several orbits around the centre and have been
tidally stripped, as evidenced by the extended $1'.3\,\times\,1'$
diffuse light around the core.

\cite{car10} estimate the velocity dispersion of the cluster to be
$1142\pm125$ km~s$^{-1}$, making Abell 3827 a rather massive
cluster. The relative radial velocities of $N1-N3$ are within 400
km~s$^{-1}$ of each other. The velocity of $N4$ is about $-1000$
km~s$^{-1}$ compared to the systemic velocity.  $N5$ is probably not a
part of the cluster core because its velocity is about $+4000$
km~s$^{-1}$ and its magnitude is $\sim 2$ fainter compared to that of
the other four ellipticals.

The cluster core lenses (at least) two background sources; the main
source, at $z=0.204$, is extended and consists of at least
three knots, $A_1,A_2,A_3$. All three appear to be lensed into quads,
but in one quad only three images are identified, and in another, only
two. The second source, at $z=0.408$, forms a single arc; a
counter arc may be lost in in the glare of the ellipticals. Based
on their spatial distribution, the images of the quads can be labelled
by their time order, even though time delays have not been measured
\citep[cf.][]{sw03}.  Accordingly, we will write $A_{ij}$ to mean the
$j$th arriving image of the $i$th knot within source $A$.  The images
to the North-East of the cluster's centre of light are the 1st
arriving ones; the two sets of images in the West are (clockwise) the
3rd and 2nd arriving, and finally the images to the South-East are 4th
arriving. The 2nd and 3rd arriving images are close together and form
a nearly merging pairs across the critical curve.

\section{Mass reconstructions}

\subsection{Free-form vs. fixed-form}

One can loosely divide the lens modeling methods into two categories,
depending on the prior information they use. The fixed-form methods
assume that mass follows light, and parametrize this relation with a
handful of scalings. Free-form methods use different types of priors;
most notably they do not require that mass trace light.

The authors of the discovery paper, \cite{car10} carried out a lensing
mass reconstruction using LENSTOOL, a fixed-form method
\citep{j07}. The reconstruction reproduces the positions of only 5 of
the 10 images well. The mean scatter between observed and recovered
images is less than $0''.8$, while the astrometric uncertainty is
$0''.2$. Overall, the mass model has $\chi^2\sim 17$ per degree of
freedom, indicating that there is additional information in the
astrometry.

This is not surprising; the dynamical state of the cluster core is
probably unsettled, so it is quite possible that light does not follow
mass. This, combined with fact that Abell 3827 has a relatively large
number of lensed images---ten---makes a free-form reconstruction a
logical choice.

%The main objection against free-form reconstructions is that they do
%not make use of the prior knowledge about galaxies and clusters, and
%hence the recovered mass ensembles include---or are even dominated
%by---mass maps that bear no resemblance to anything that nature can
%come up with. Of course a complimentary criticism can be levied
%against fixed-form methods, which underestimate the diversity of mass
%distributions that are found in nature, especially in non-isolated
%systems.

We use a free-form reconstruction method, PixeLens
\citep{sw04,2008ApJ...679...17C}.  The reconstruction is done in a
circular window, which is divided into many equal-sized square
pixels. The mass within each pixel is uniformly distributed but each
pixel's mass value is allowed to vary independently of
others. PixeLens takes advantage of the fact that the lensing equation
is linear in the mass pixels and in the coordinates of the source, and
so it is relatively straightforward to come up with solutions that
reproduce image positions exactly. In fact, since the unknowns
outnumber the knowns (coordinates of the lensed images) by a large
factor, an infinity of solutions are possible. PixeLens generates a
user-specified number of individual models, according to a weak and
adjustable prior of which the ensemble average is considered in this
paper. The prior and the model-sampling technique are motivated and
explained in detail in the above papers, so here we just remark on two
points that are important for the modelling.

First, along with the image positions, a notional centre must be given.  
The default behaviour of PixeLens is to require the local
density everywhere to point at most $45^\circ$ away from the centre.
In this work we increase the allowed angle (called {\tt dgcone} for
``density-gradient cone'') somewhat, typically to $65^\circ$.  This
allows more substructure, as appears to be demanded by the data.

Second, to account for the mass exterior to the modeling window,
PixeLens can add external shear of any magnitude. The shear axis can
be within $45^\circ$ of a user-specified shear direction, denoted by
{\tt shear}.  We typically set {\tt shear 60} ---on the basis of the
axis of the diffuse light on arcminute scale around the cluster
core--- which permits any shear orientation from $15^\circ$ to
$105^\circ$.

The version of the lens-reconstruction software used is included as an
online supplement.  Other examples of free-form techniques for cluster
lenses are described in \cite{2007MNRAS.380.1729L} and
\cite{2008ApJ...681..814C}.

\subsection{Fiducial reconstruction}\label{fidu}

To start with, we create a reference, or fiducial reconstruction. It
uses all the 10 images identified by \cite{car10}. As the centre we
use the centre of light (also the centre of their Figure 2 frame),
which is about $2''$ West of N2.

At the redshift of the cluster, $z=0.1$, the critical surface mass density
for sources at infinity is $\Sigma_{\rm crit}=0.953$ gm/cm$^2$, and
$1''$ corresponds to 1.93 kpc, for $\Omega_m=0.3$,
$\Omega_\Lambda=0.7$, and $h=0.67$. The diameter of the reconstruction
window was set to 29 pixels, with the scale of $1.43''$/pixel, so the
angular and physical size of the window are about $40''$ and 77 kpc,
respectively.

\begin{figure}%[t]
%\epsscale{1.5}
%\plotone{v192/Casek_mass.eps}
%\plotone{smfiducialmass.eps}
\plotone{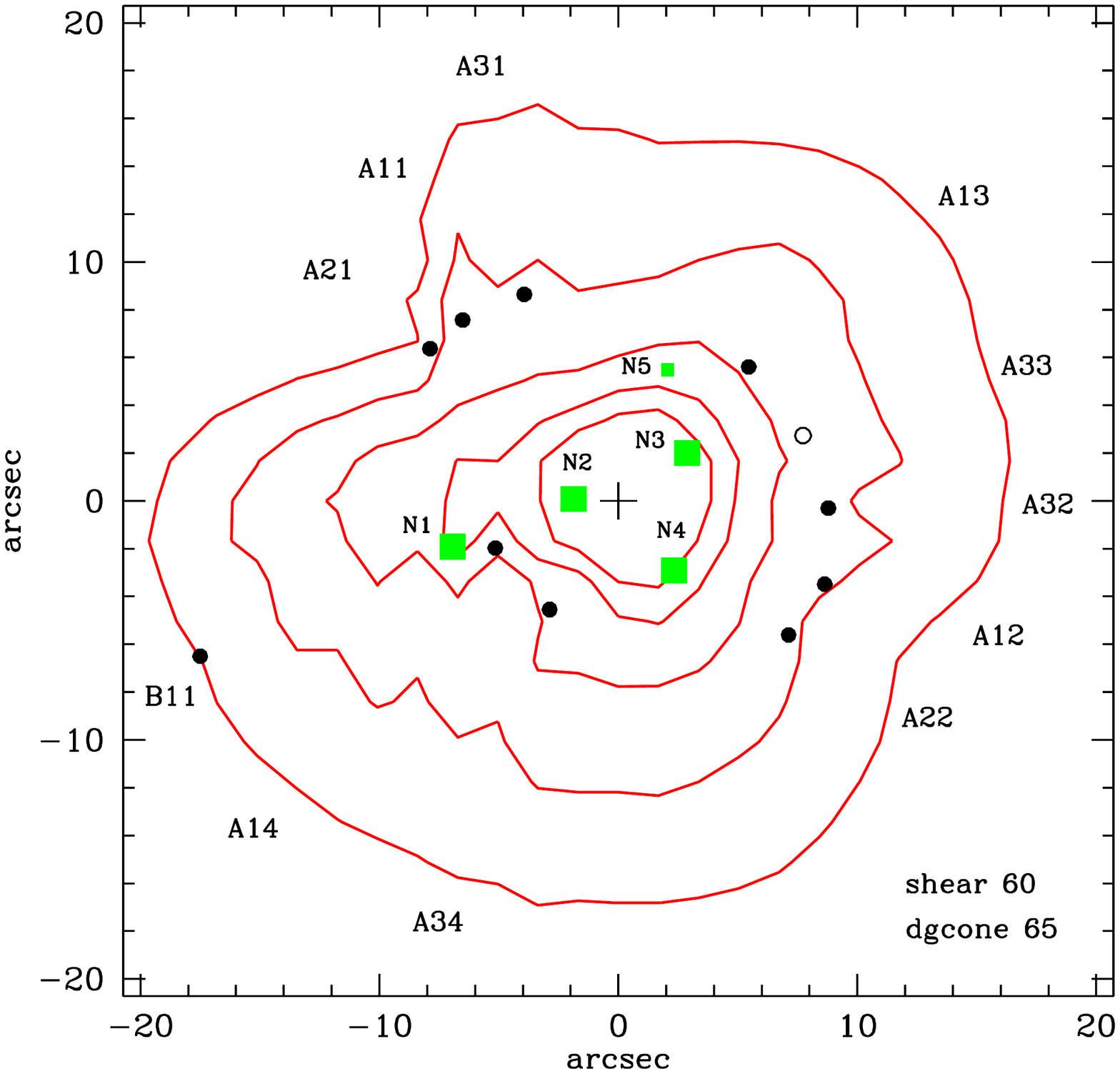}
\vskip-1.5in
\caption{Fiducial mass reconstruction. The map is an average of 500 individual mass maps. 
Contours of equal projected mass density are at 0.9, 1.3, 1.7, 2.1 and 2.5  of $\Sigma_{\rm crit}$, 
where $\Sigma_{\rm crit}$ is the critical density for $z_l=0.1$ and sources at infinity.
Images identified by \cite{car10} are marked with black solid dots, and labeled at the periphery 
of the figure. $A_{ij}$ means the $j$th arriving image of the $i$th knot within source $A$. 
The empty black circle is the $A_{33}$ image that the present authors identified from the observed 
image. It is not used in the fiducial but it used in some later reconstructions. Green squares 
are ellipticals; N5 is marked with a smaller symbol because it is probably not a part of the 
cluster core. At the redshift of the cluster, $1''=1.93\;h^{-1}_{0.67}$~kpc.} 
\label{kmass}
\epsscale{1}
\end{figure}

\begin{figure}%[t]
%\plotone{v192/smmassenc.pdf}
\plotone{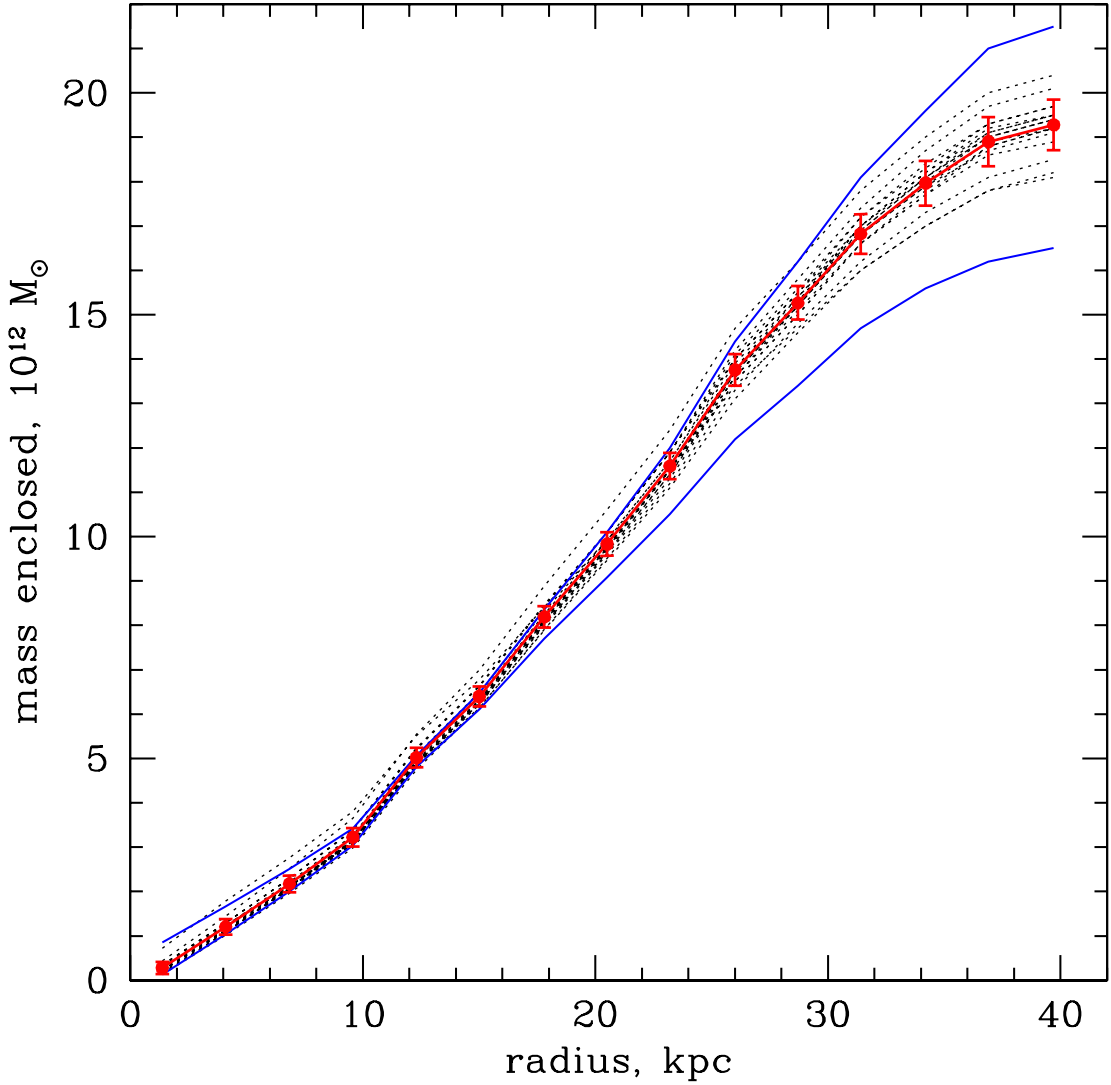}
\vskip-1.5in
\caption{Mass enclosed as a function of distance from centre. The central thick red line
with points and errorbars represents the average and rms dispersions of 18 reconstructions
shown in Sections~\ref{effp} and \ref{effs}. The think black dotted lines are the 18 
individual reconstructions. Our fiducial reconstruction is indistinguishable from the average. 
The two blue solid flanking lines enclose 90\% of the 500 individual mass models of the fiducial 
reconstruction.}
\label{massenc}
\epsscale{1}
\vskip0.0in
\end{figure}

The fiducial PixeLens reconstruction is shown in Figure~\ref{kmass};
the mass density contours are spaced by $0.5\,\Sigma_{\rm crit}$. The
map is an average of 500 individual models. (The mass map is not
sensitive to the number of individual models in the ensemble as long
as it is greater than $\sim 100-200$, the pixel size within the range
$1.25''$-$2''$ per pixel, or the orientation of the reconstruction
window with respect to the cardinal directions.)  Figure~\ref{massenc}
is a plot of mass enclosed as a function of radius. Our fiducial
reconstruction is indistinguishable from the average of 18
reconstructions discussed in Sections~\ref{effp} and \ref{effs}, and
shown as the red thick line. The thin dotted lines are all the 18
models, and the errorbars represent their rms dispersion. The two blue
solid flanking lines enclose 90\% of the 500 individual mass models of
the fiducial reconstruction. Note that at large radii, i.e. outside the
images, the enclosed mass begins to level off since the image
positions do not require any mass at those radii. However, the 90\%
errorbars are fully consistent with the density profile continuing on
with the same slope as at smaller radii.

\begin{figure}%[t]
\epsscale{1.5}
\vskip-5in
%\plottwo{v192/Casek_arriv1.eps}{v192/Casek_arriv3.eps}
%\plottwo{fig2a.pdf}{fig2b.pdf}
\plottwo{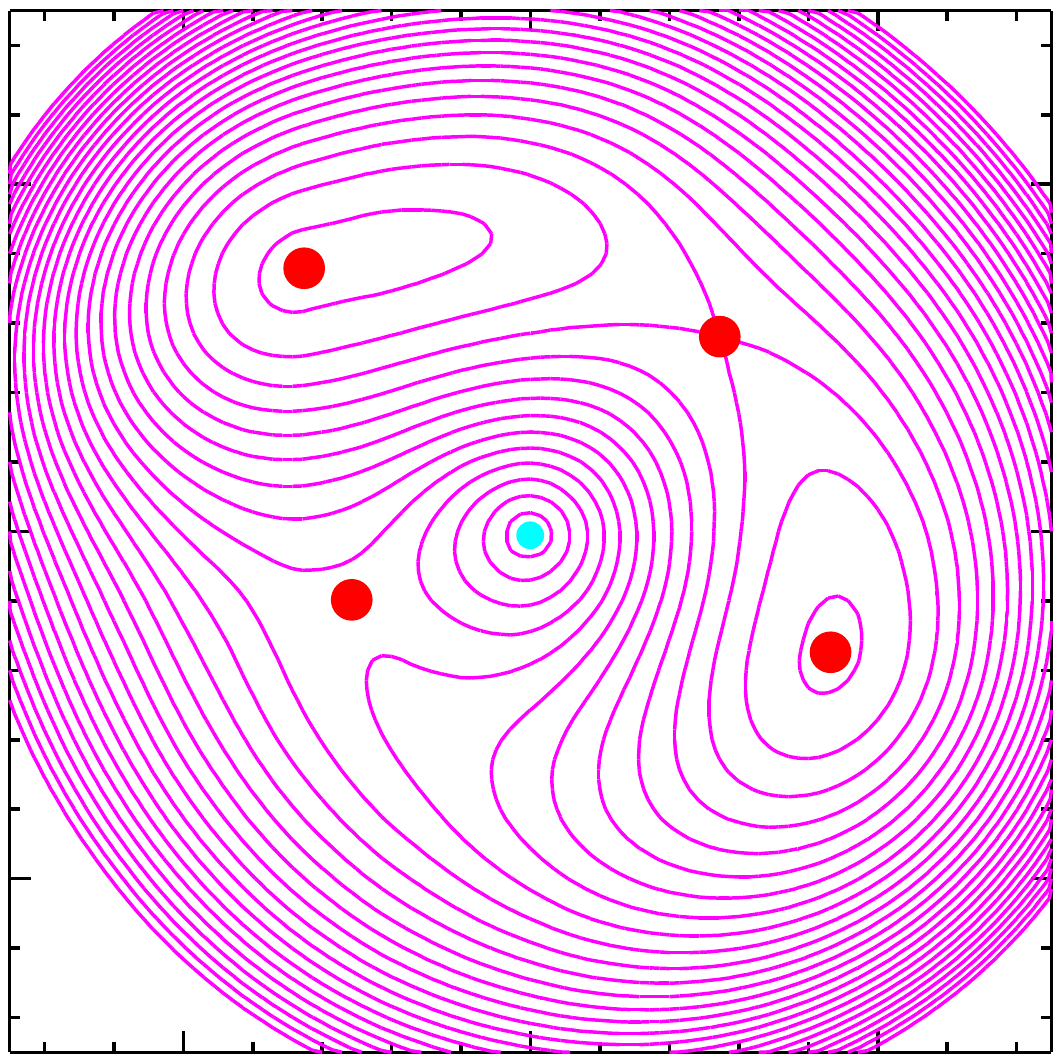}{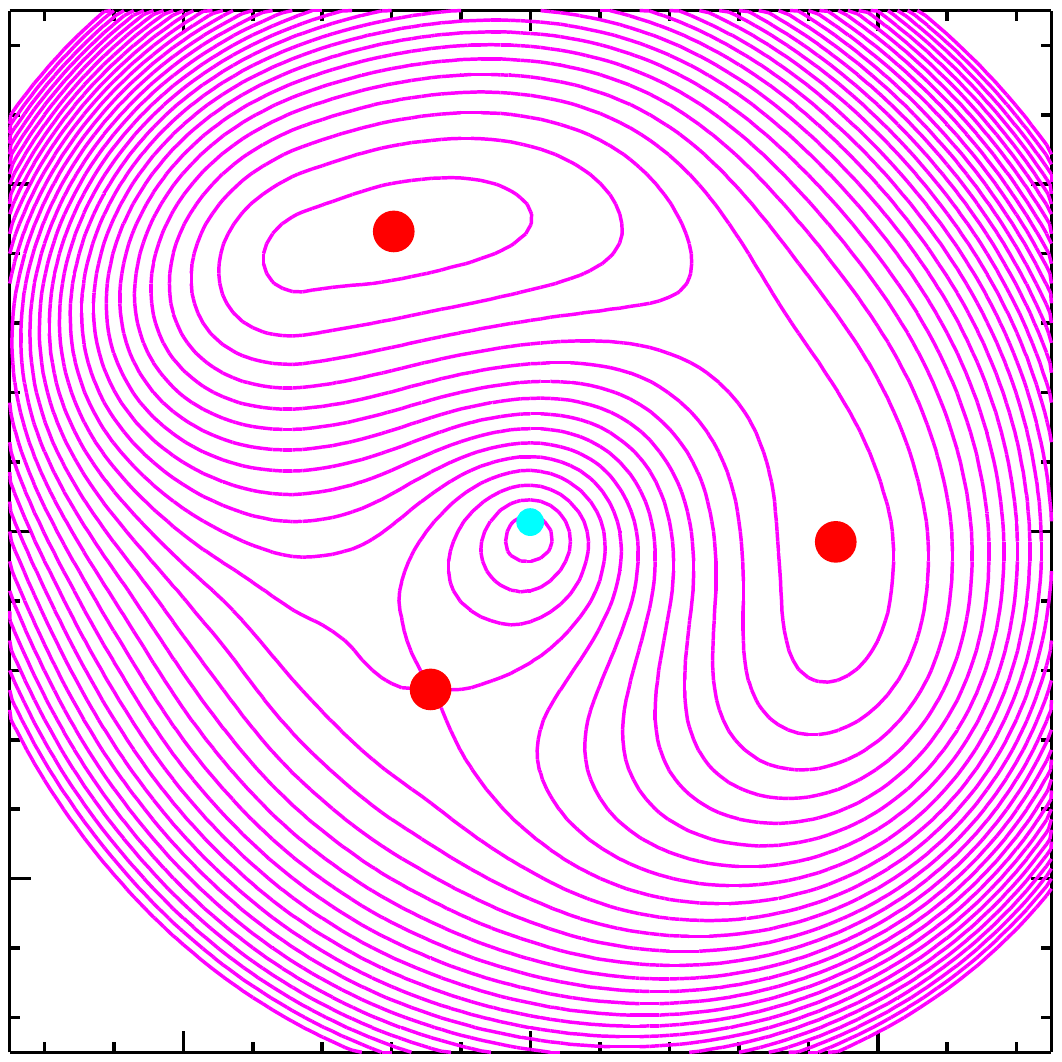}
\vskip+0.5in
\caption{Arrival time surfaces of PixeLens fiducial reconstructions of sources $A_1$ and $A_3$.}
\label{karriv}
\epsscale{1}
\vskip0.0in
\end{figure}

Figure~\ref{karriv} shows the arrival time contours of the $A_1$ and
$A_3$ knots of source $A$.  From the contours, we see that the model
indicates extra images appearing next to the first arriving image of
$A_1$.  These may be spurious, or may actually be present, depending
on the local mass substructure.  They do not, however, affect the
larger-scale features of the mass maps, which are the focus of this
paper.  The plot of arrival-time contours also provides a simple model
for the ring-like extended image \citep[cf.][]{sw01} which in this
case arises from the extended source $A$.  Detailed comparison would,
however, be very difficult because the ring is faint and in many
places overlaid with the light from the ellipticals.

We have not attempted to use the arc $B$ as a lensing constraint,
since it has no identified counter image. In principle it is possible
to extract some strong-lensing information from this arc.  If
pointlike features could be identified, they could be put in as
multiple-image constraints in the usual way.  Another possibility
would be to assume two fictitious image positions along the arc, which
in models would have the effect of forcing a critical curve to pass
through the arc.

\begin{figure}%[t]
%\plotone{v192/smprofile.pdf}
\plotone{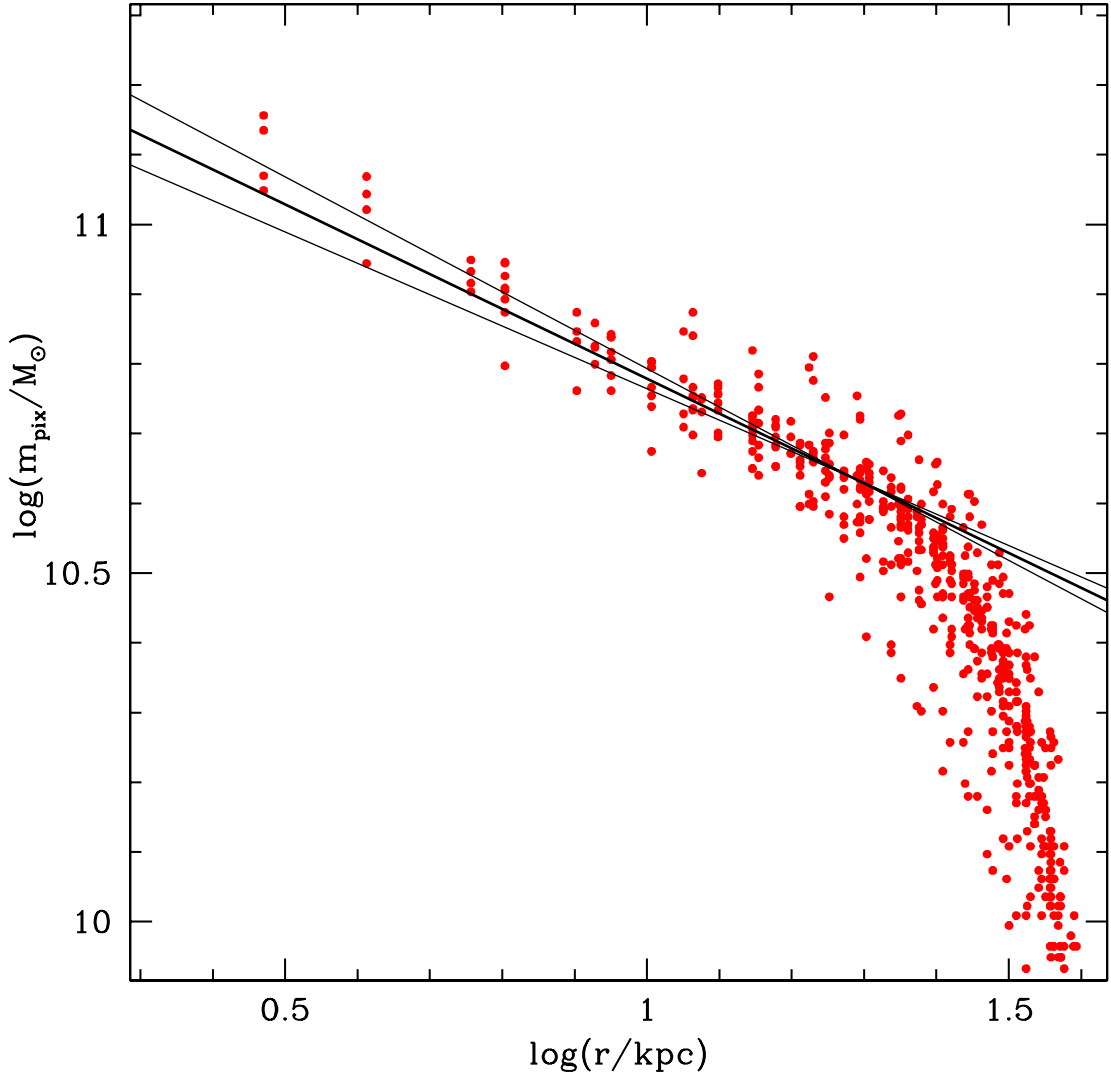}
\vskip-1.5in
\caption{Pixel mass vs.\ distance from centre for the fiducial
  reconstruction --- each dot represents a mass pixel. The outermost
  multiple image is at a projected radius of 20~kpc.  (The arc $B_{11}$ is
  at 37~kpc, but no counter-images have been identified.)  There is
  thus no new strong-lensing information beyond 20~kpc, and PixeLens
  puts very little mass into pixels further out.  The three straight
  lines have log-log slopes of -0.45, -0.5 and -0.55. To reveal the
  mass substructure, or excess above the smooth cluster profile, the
  slope of -0.5 was subtracted from PixeLens ensemble average maps;
  the normalizations were chosen as described in Section~\ref{fidu}.}
\label{profile}
\epsscale{1}
\vskip0.0in
\end{figure}

The mass contours in Figure~\ref{kmass} show deviations from circular
symmetry, which are due to mass inhomogeneities in the lens. We
isolate these by subtracting a smooth component, as
follows. Figure~\ref{profile} shows pixel mass as a function of the
distance from the centre of the fiducial reconstruction. As already
mentioned when discussing Figure~\ref{massenc}, the lack of lensing
constraints beyond about 20~kpc leads PixeLens to put very little mass
into those pixels; so the tapering off of points in
Figure~\ref{profile} is an artifact. The average projected density
profile slope where the images constrain the mass well, i.e. around
$\sim 6''-24''$, is $d\ln\Sigma(R)/d\ln R\approx -0.5$, and is
representative of most mass maps in this paper. This is shown as the
thick straight line in this log-log plot; the other two lines have
slopes of $-0.45$ and $-0.55$, and represent the reasonable range of
possible power-law slopes. To isolate the mass inhomogeneities we
subtract circularly symmetric density profile of slope $-0.5$ from the
reconstructed mass maps; a change of $\pm 0.05$ in the assumed smooth
profile slope would not have made a significant difference. The
normalization of the subtracted profile is adjusted such that the
contours of substructure, or the excess mass density left
after the subtraction of the smooth cluster profile, contain 0.5, 1,
2, 4 and 8 \% of the total mass within the full $40''$ window.  This
normalization is calculated separately for each mass map and each of
the five contour levels.

\begin{figure}%[t]
%\epsscale{0.95}
%\plotone{smfiducial.eps}
%\plotone{fig3.pdf}
\plotone{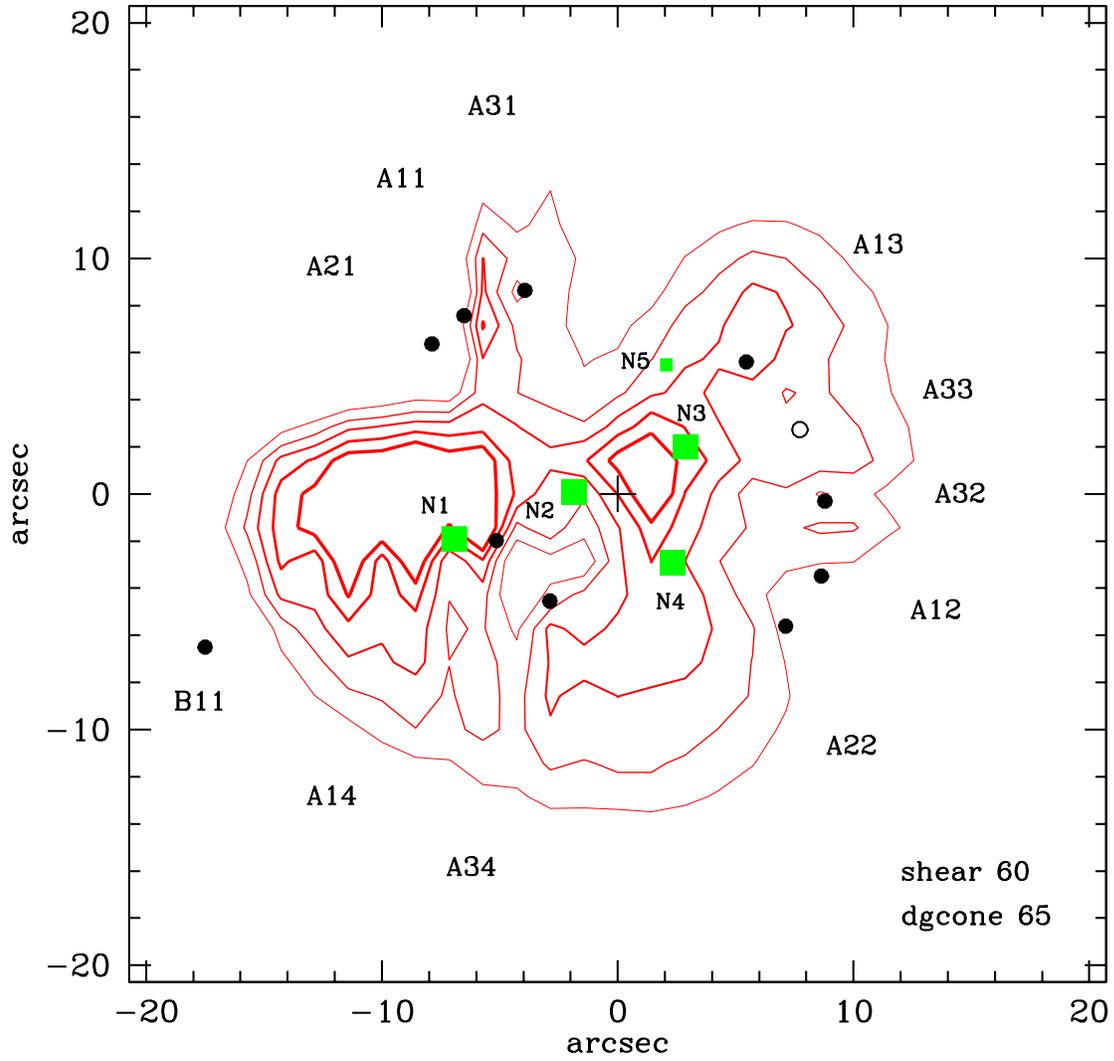}
\vskip-1.5in
\caption{Mass substructure, or excess left after subtracting the smooth cluster profile from the 
fiducial mass reconstruction shown in Figure~\ref{kmass}. The 5 contours, from thin to 
thick ones contain 8, 4, 2, 1, and 0.5\% of the total reconstructed mass. The rest of the
annotations are the same as in Figure~\ref{kmass}.} 
\label{fiducial}
\end{figure}

Figure~\ref{fiducial} shows the five contour levels of the mass
substructure, as red solid lines; thicker lines represent contours of
smaller enclosed mass. Images are shown as black solid dots, and
labeled at the periphery of the figure. The arc $B_{11}$ has no
detected counter images.  Image $A_{33}$ was not identified by
\cite{car10}, but looking at their Figure 2, the present authors feel
$A_{33}$ is the location of that image. It is not used in the fiducial
reconstruction but is used in some other ones. Galaxies are denoted by
green squares and labeled as in \cite{car10}; $N5$ is a smaller square
because the galaxy is probably background and less massive.

Figure~\ref{fiducial} shows that the mass substructure does not trace
the visible galaxies, an indication that the cluster core is not in
equilibrium, but is dynamically disturbed.  There is one dominant mass
clump, to the North East (upper left) of galaxy $N1$, but not centered
on $N1$. This clump is the main subject of the paper, and we will
refer to it as the NE-$N1$ clump. A smaller clump is between galaxies
$N3-N4$, and seems to be avoiding $N2$.  The following sections will
demonstrate that even though the details of the reconstructions
differ, the main features of this fiducial mass map, and most
importantly the existence and location of NE-$N1$ clump, and the lack
of mass associated with $N2$, are robust to changes in priors, centre
of the reconstruction window, and, to some degree, the subsets of
images used in reconstruction.

\subsection{Effect of priors used in reconstruction}\label{effp}

Figure~\ref{lots3} shows maps of substructure mass contours for nine
reconstructions. The image set used is the same as in the fiducial
case, but the priors are different.  The assumed values of {\tt
dgcone}, and {\tt shear} are shown in each panel. Since the cluster
core is most likely not in equilibrium, we tried three different lens
centres.  The fiducial one is called $C0$, and was identified by
\cite{car10} as the cluster's centre of light; it sits between $N2$
and $N3$. Centre $C1$ is shifted by $(+1''.5, +1'')$ from $C0$ and is
close to $N3$. Centre $C2$ is shifted by $(-1''.875 +0''.08)$ with
respect to $C0$; it coincides with the location of galaxy $N2$, the
central-most elliptical in the core. The three choices for the lens
centre are used in the reconstructions of the three rows of
Figure~\ref{lots3}, respectively. Their location is marked with a
cross.  (We remark again that the lens centre in PixeLens is needed
because the density gradients, used for {\tt dgcone}, are calculated
with respect to the lens centre. The other utility of the lens centre
will be put to use in Section~\ref{cent}.)

Figure~\ref{lots3} shows that all reconstructions, with the possible
exception of the middle panel of the third row, require a mass lump to
the NE of galaxy $N1$. The size and position of the lump vary between
panels, but in none of these is the lump centered on $N1$; it is always
to the upper left of that galaxy. The other common feature is that
$N2$ has relatively little excess mass associated with it. The nearby
mass clump sometimes encompasses $N3$ or $N4$, but not $N2$.

\subsection{Effect of image subsets used in reconstruction}\label{effs}

Figure~\ref{lots4} shows the effect of leaving out some of the lensed
images from the reconstruction. The images plotted in each panel are
the only ones used in the corresponding reconstruction. The lens
centre is always $C0$, and the PixeLens priors are shown in each
panel. Most changes to the image set makes little difference to the
global features of the recovered mass maps, and specifically the
NE-$N1$ mass clump.

Since this mass clump is of primary interest for us, we examine its
dependence on the images more closely. From this Figure, and other
reconstructions not shown here, we know that the single most important
image that determines the location of the clump near $N1$ is image
$A_{14}$. In the lower right panel $A_{14}$ was not used, and the mass
clump moved to a different, lower, or more Southerly, location. In
that reconstruction, both the mass distribution and the image
configuration have an approximate bilateral symmetry, with the axis
inclined by $\sim 60^\circ$ to the $+x$-axis. The location of $A_{14}$
predicted by such a bilaterally symmetric mass distribution would have
been very close to the observed $A_{34}$. However, the observed
$A_{14}$ is about $4''$ clockwise from $A_{34}$.

To make that happen, one needs a large mass clump roughly where
PixeLens puts the NE-$N1$ clump. This can be understood in terms of
the arrival time surface. $A_{14}$ is a saddle point image. A
sufficiently large mass just outside of the image circle will create a
local bump in the arrival time surface and a saddle point will be
created between it and the centre of the lens.

In Section~\ref{expl} we will present additional arguments to show why
$A_{14}$ requires a large mass to exist beyond the location of galaxy
$N1$.

%This lens reconstruction looks like 1422. The axis of bilateral
%symmetry is inclined by $60^\circ$, and the 3 quads are displaced
%along that axis in the order from down most to top most: A2, A1, A3,
%and images \#4 are on that axis, and very close together.

One might be tempted to say that $A_{14}$ has been misidentified, and
is actually much closer to $A_{34}$. If that is assumed, the
reconstructed mass and observed galaxy positions bear no
correspondence to each other, as shown in the lower right panel of
Figure~\ref{lots4}. Furthermore, the imaging data as presented in
\cite{car10} shows a definite, though faint smudge at the location of
$A_{14}$. We conclude that $A_{14}$ is real, and hence the true mass
distribution is close to what our fiducial reconstruction looks like.

\subsection{Properties of the NE-$N1$ mass clump}\label{props}

The mass substructure, i.e. the excess left after subtracting the
smooth cluster profile, of all eighteen reconstructions of
Figures~\ref{lots3} and \ref{lots4} are superimposed in
Figure~\ref{superpose}. For clarity, we plot only the contours
enclosing 1\% of the total mass in every reconstruction. The figure
highlights the features common to most reconstructions: the massive
NE-$N1$ mass clump, and the secondary clump around galaxies $N3$ and
$N4$, which avoids $N2$.

The main goal of the paper is to estimate the lower limit on the dark
matter cross-section using eq.~\ref{dispunits} as applied to galaxy
$N1$ and the NE-$N1$ mass clump. In this section we estimate the
parameters associated with the NE-$N1$ clump, namely, its mass,
$M_{dm}$, size, $s$, and distance from $N1$, $d$, and give a measure
of dispersion in these properties between reconstructions.

Figure~\ref{superpose} gives a visual impression of the systematic
uncertainties in our reconstructions. Because the shape and location
of the NE-$N1$ clump is not exactly the same in various
reconstructions, there is no unique way of quantifying its average
properties and estimating the dispersion in these properties between
the various mass maps.

We chose to use an iso-density mass substructure contour to define the
clump, and the corresponding mass excess within that contour as the
clump's mass, its center of mass as the center of the clump, and the
enclosed area, $A$, to define the size of the clump,
i.e. $s=(A/\pi)^{1/2}$.  First, we need to decide what to take as the
boundary of the clump.  In most reconstructions, including the
fiducial one, the mass substructure contours containing 0.5 and 1\% of
the total mass are disjointed, and enclose separate mass clumps, while
the contours containing 4 and 8\% of the total mass form one
contiguous region. This suggests that $\sim 1\%$ of the total
recovered mass is in the individual clumps, therefore we take the
closed 1\% contour located to the NE of the $N1$ galaxy as defining
the clump.

Two of the reconstructions, the bottom middle in Figure~\ref{lots3}
and the bottom right in Figure~\ref{lots4} have no excess mass at 1\%
level in the vicinity of $N1$, so we exclude them.  With these
choices, the average and rms dispersion in the mass of the NE-$N1$
clump based on the 16 maps is $M_{dm}=(1.22\pm 0.44)\;
10^{11}M_\odot$, its distance from the centre of the elliptical $N1$
is $d=2.68''\pm 1.08''$, and its size $s=4.28'' \pm 2.46''$.  Our
fiducial reconstruction gives $M_{dm}=1.47\; 10^{11} M_\odot$,
$d=2.50''$, and $s=4.11''$, all of which are within the rms of the 16
maps, so when estimating $\sigma/m$ in Section~\ref{dmx} we will use
the fiducial values.

We also need to estimate $r$, the clumps distance from the cluster
centre.  Equation~\ref{disp} assumes that the centre of the elliptical
galaxy and the centre of its displaced dark matter halo are
equidistant from the cluster centre, i.e. $r_{st}\approx r_{dm}$ since
the displacement $d$ is small. Therefore we take $r=r_{st}\approx
7.2''$, same for all reconstructions. The uncertainty in $r$ comes
from the uncertainty in the location of the cluster centre. In
Section~\ref{cent} below we show that the latter is of order
$1''-2''$, which translated into a fractional error, which is smaller
than that in the other two relevant lengths, $d$ or $s$.

While the errors in the estimated properties of the NE-$N1$ mass clump
are relatively small, $\simlt 50\%$, we stress that these are not the
main sources of uncertainty in estimating the parameters of the dark
matter particle cross-section, $\sigma/m$. The dominant sources of
uncertainty are the unknown dynamics of the cluster, the evolution in
cluster core's density profile slope, and the galaxy orbit shape,
especially that of elliptical $N1$ (see Section~\ref{intro}), and
cannot be quantified using lensing data. To properly address these one
would need to carry out a series of N-body simulations with a range of
initial conditions.

\subsection{Effect of image position uncertainties}\label{uncert}

Based on their observations, \cite{car10} quote image astrometric
uncertainty of $0''.2$.  Since PixeLens reproduces image positions
exactly, we need to check if astrometric errors could have affected
our results. Here we redo the fiducial reconstruction but randomly
shift the observed images within $\pm 0.5''$ of their true positions.
Five realizations are shown in Figure~\ref{fake}. To reduce run-time,
we used 300 models to create an ensemble, instead of the usual
500. Since all five reconstructions look very similar and have small
rms dispersions ($M_{dm}=(1.39\pm 0.23)\; 10^{11}M_\odot$ and
$d=2.77''\pm 0.17''$), smaller than the ones presented in the
Section~\ref{props} we conclude that astrometric errors have not
affected our results.

\subsection{Centre of mass of the cluster}\label{cent}

\begin{figure}%[t]
%\epsscale{0.95}
%\plotone{smlots3.eps}
%\plotone{v192/smlots3mass.pdf}
\plotone{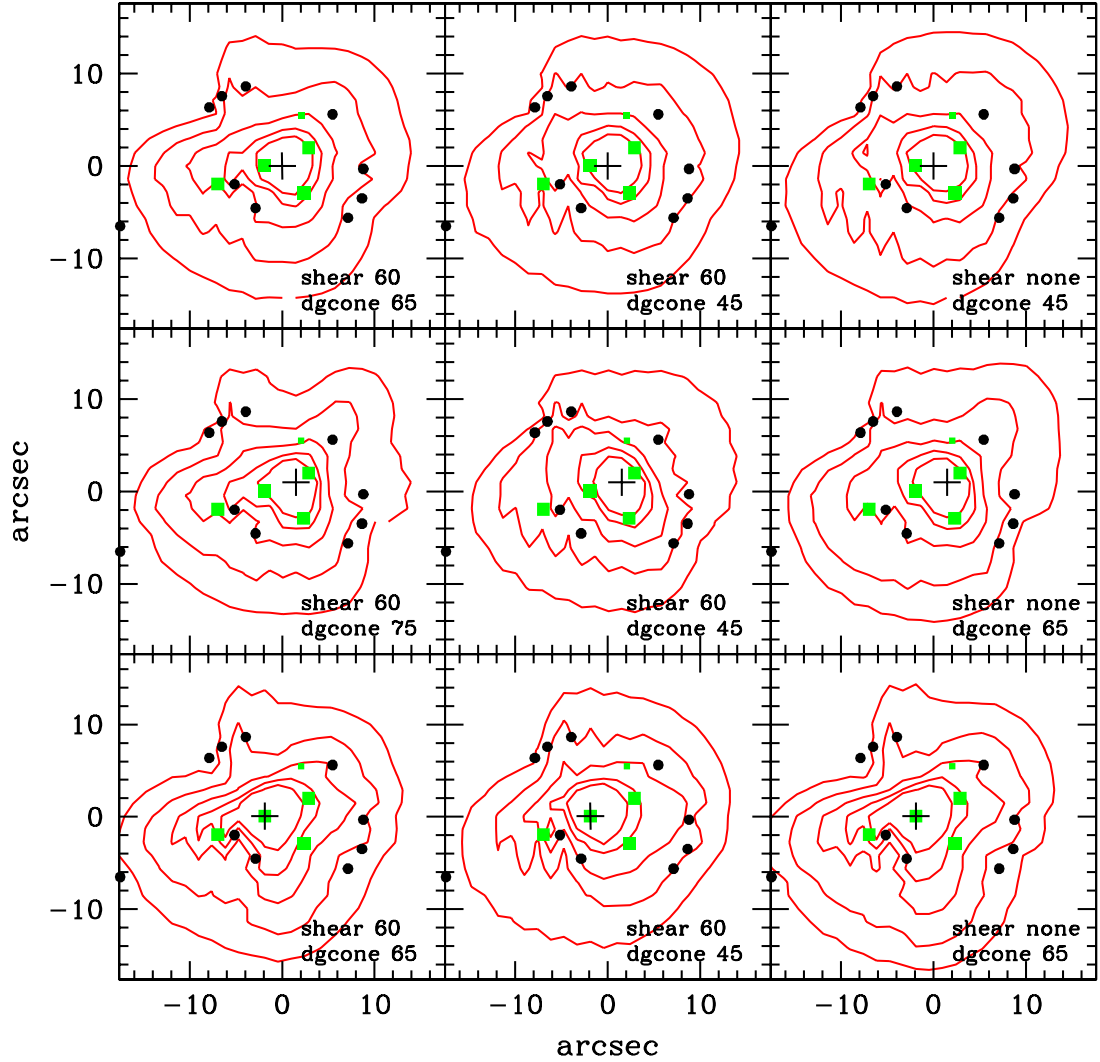}
\vskip-1.5in
\caption{PixeLens mass reconstructions using different priors and lens centres. The values of 
{\rm shear} and {\tt dgcone} are indicated in each panel; the centre of each reconstruction
window is marked with a cross, and is different in each of the three rows. The top row assumes
the fiducial lens centre. The whole fiducial lensed image set (10 images) is used in all nine 
reconstructions; images are marked with black solid points. The top left map is the fiducial map. 
The mass density contours are at 0.9, 1.3, 1.7, 2.1, and 2.5 of critical surface mass density 
for lensing for sources at infinity.} 
\label{lots3mass}
\end{figure}

\begin{figure}%[t]
%\epsscale{0.95}
%\plotone{smlots3.eps}
%\plotone{fig4.pdf}
\plotone{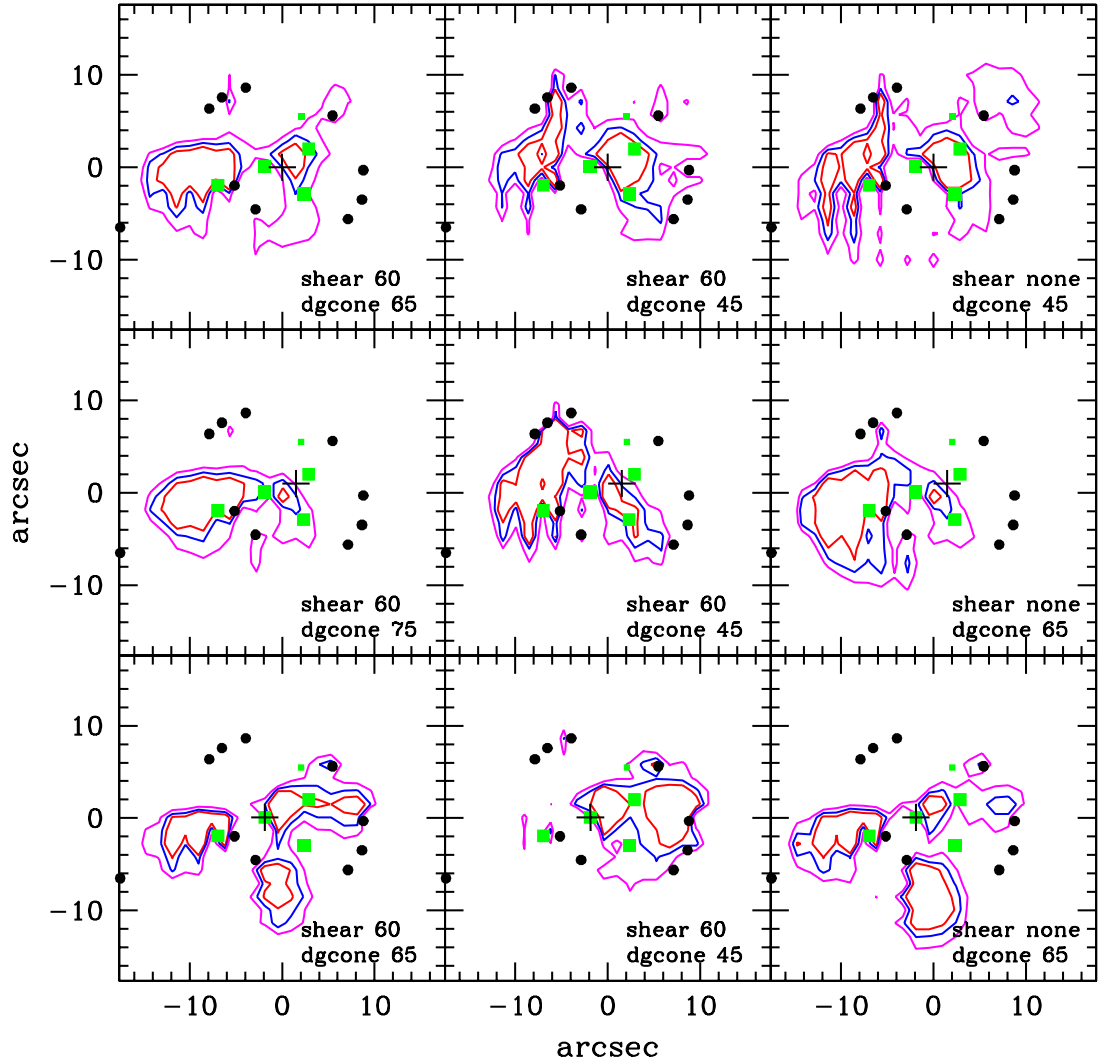}
\vskip-1.5in
\caption{Mass substructure left after subtracting smooth density profiles from PixeLens mass maps
shown in Figure~\ref{lots3mass} (see Section~\ref{fidu}). The density contours containing 
0.5, 1 and 2\% of the total reconstructed mass are plotted as red, blue and magenta curves, 
respectively.} 
\label{lots3}
\end{figure}

\begin{figure}%[t]
%\epsscale{0.95}
%\plotone{smlots3.eps}
%\plotone{v192/smlots4mass.pdf}
\plotone{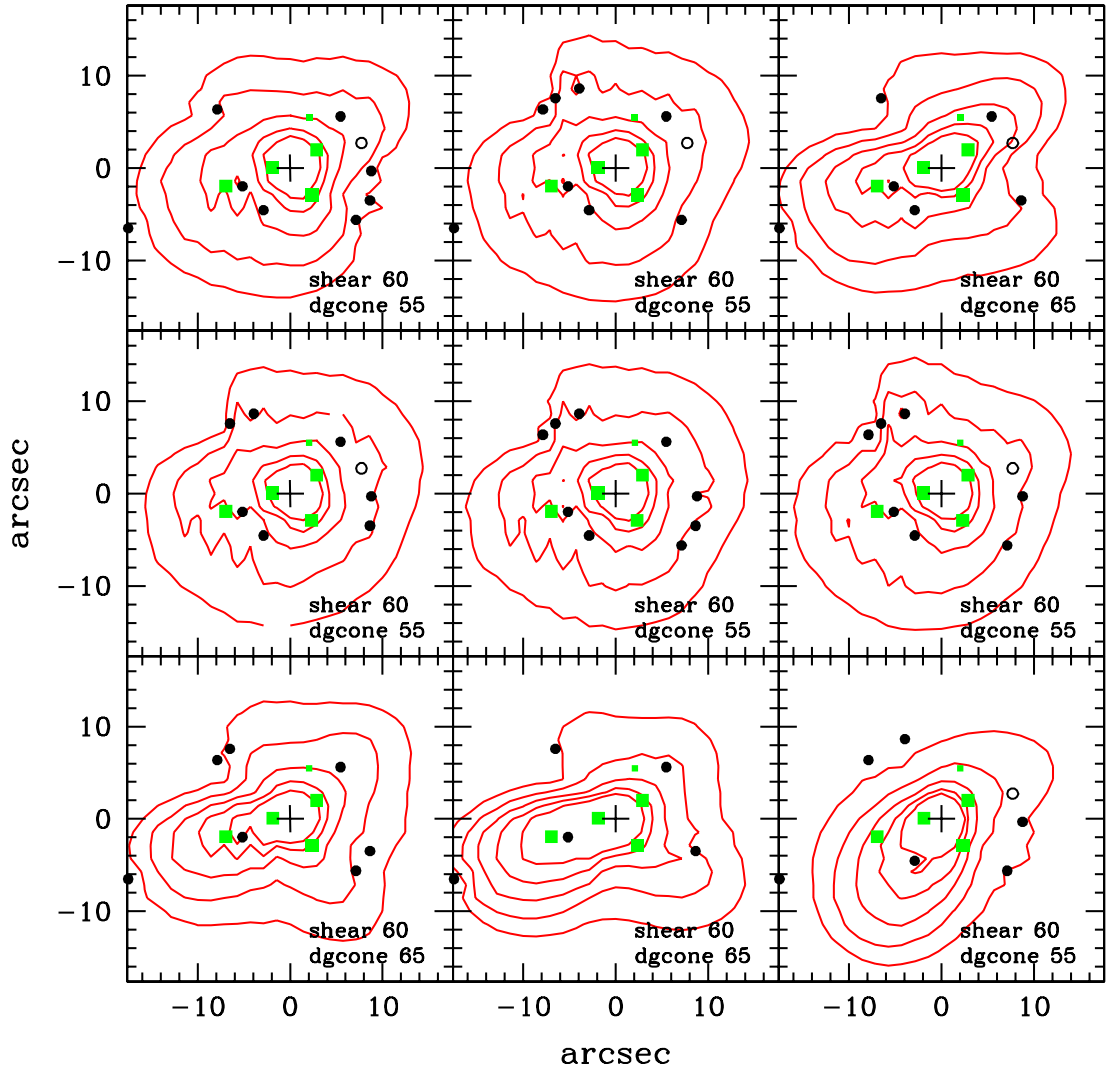}
\vskip-1.5in
\caption{Similar to Figure~\ref{lots3mass}, but using different sets of lensed images,
which are indicated in each panel. The lens centre is the same in all nine reconstruction 
(same as fiducial), and is marked with a cross.} 
\label{lots4mass}
\end{figure}

\begin{figure}%[t]
%\epsscale{0.95}
%\plotone{smlots4.eps}
%\plotone{v192/smlots4.pdf}
\plotone{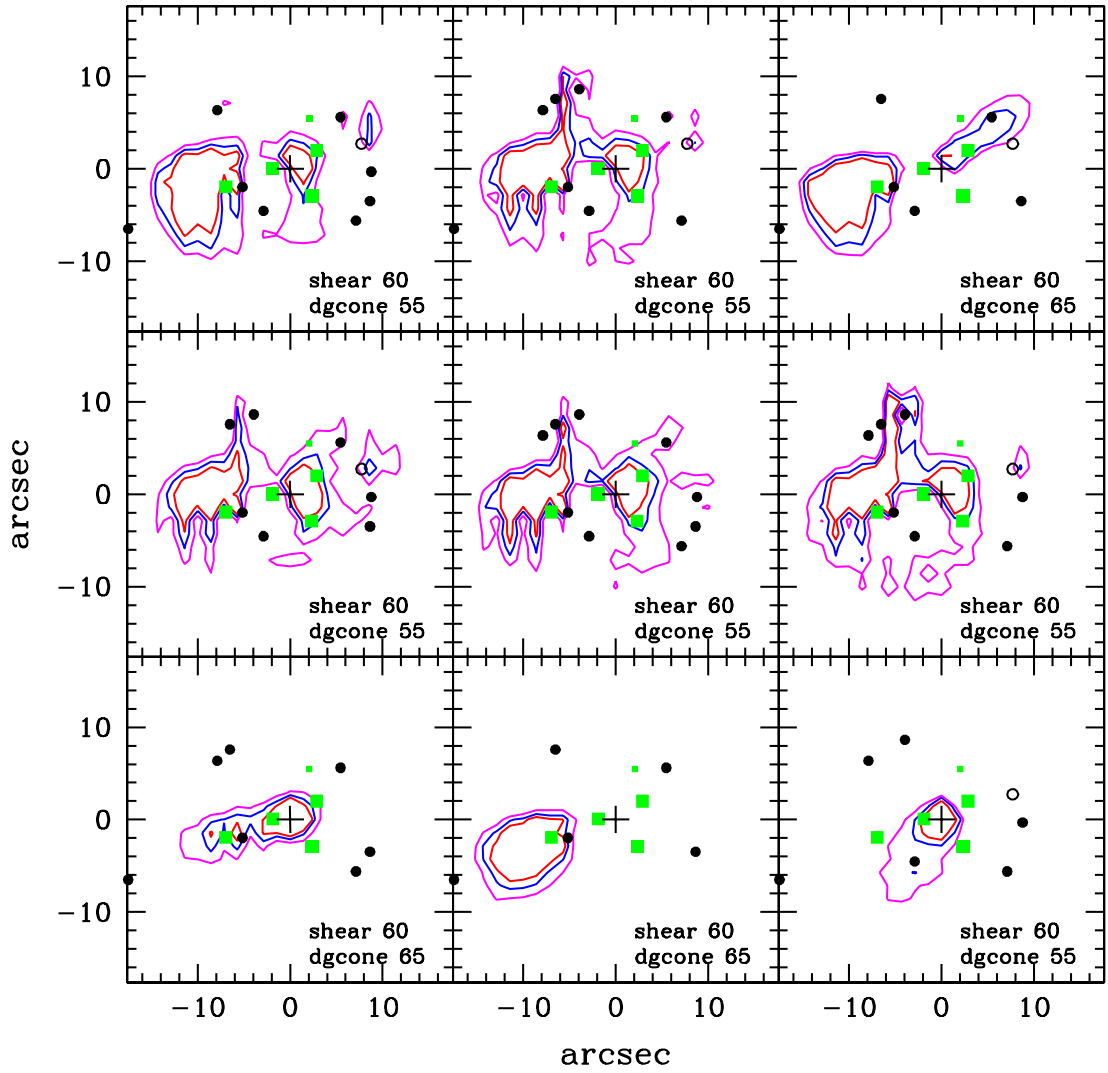}
\vskip-1.5in
\caption{Similar to Figure~\ref{lots3}, i.e. mass substructure, but for PixeLens maps shown in 
Figure~\ref{lots4mass}.} 
\label{lots4}
\end{figure}

\begin{figure}%[t]
%\epsscale{0.95}
%\plotone{smsuperpose.eps}
%\plotone{fig6.pdf}
\plotone{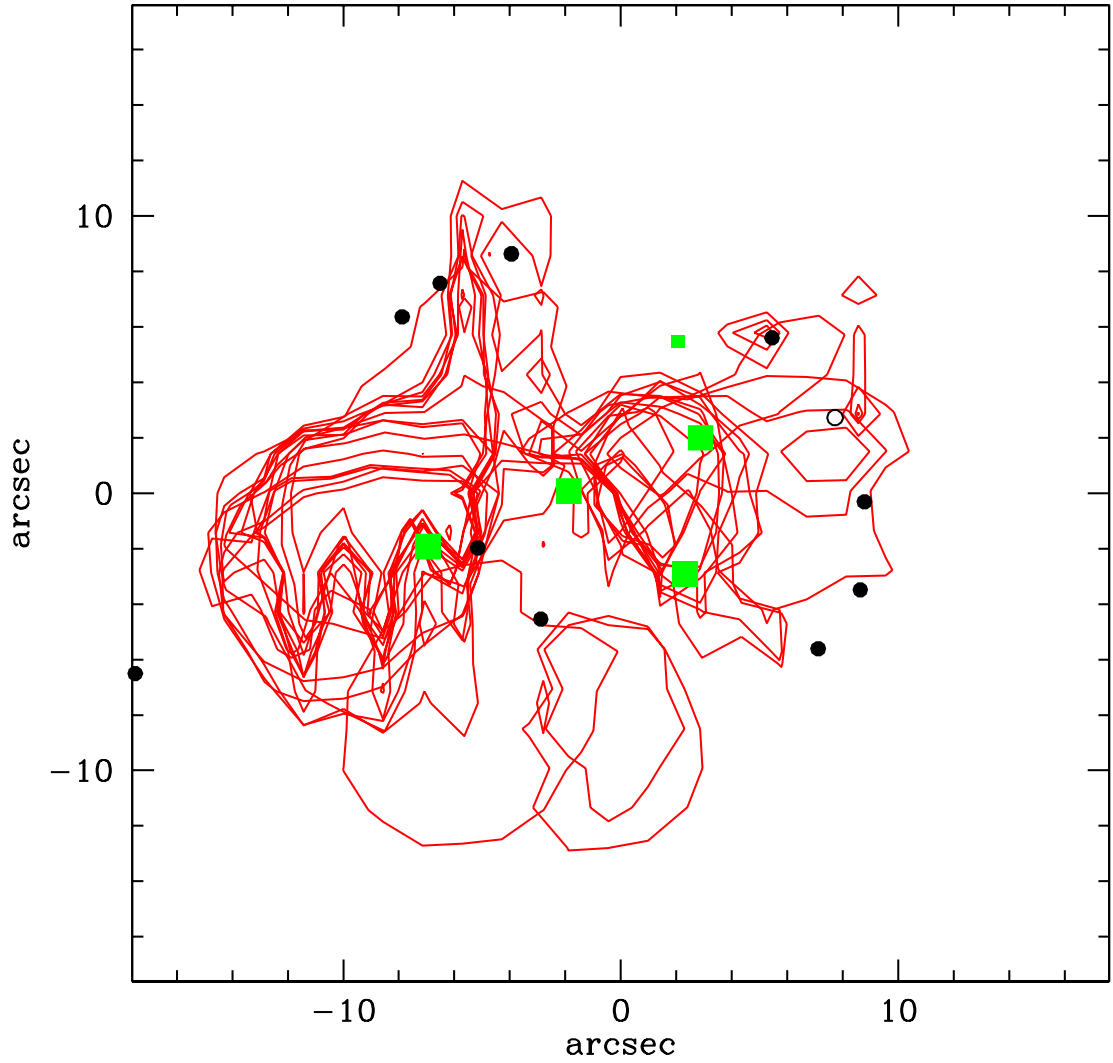}
\vskip-1.5in
\caption{All eighteen reconstructions from Figures~\ref{lots3} and \ref{lots4}, but only
one contour per reconstruction is shown, the one enclosing 1\% of the total mass. As before,
galaxies are represented by green squares, lensed images are black dots. This plot highlights
the features common to all reconstructions: the massive NE-$N1$ mass clump, and the secondary
clump around galaxies $N3$ and $N4$, which avoids $N2$.} 
\label{superpose}
\end{figure}

\begin{figure}%[t]
%\epsscale{0.95}
%\plotone{smfake.eps}
%\plotone{fig7.pdf}
\plotone{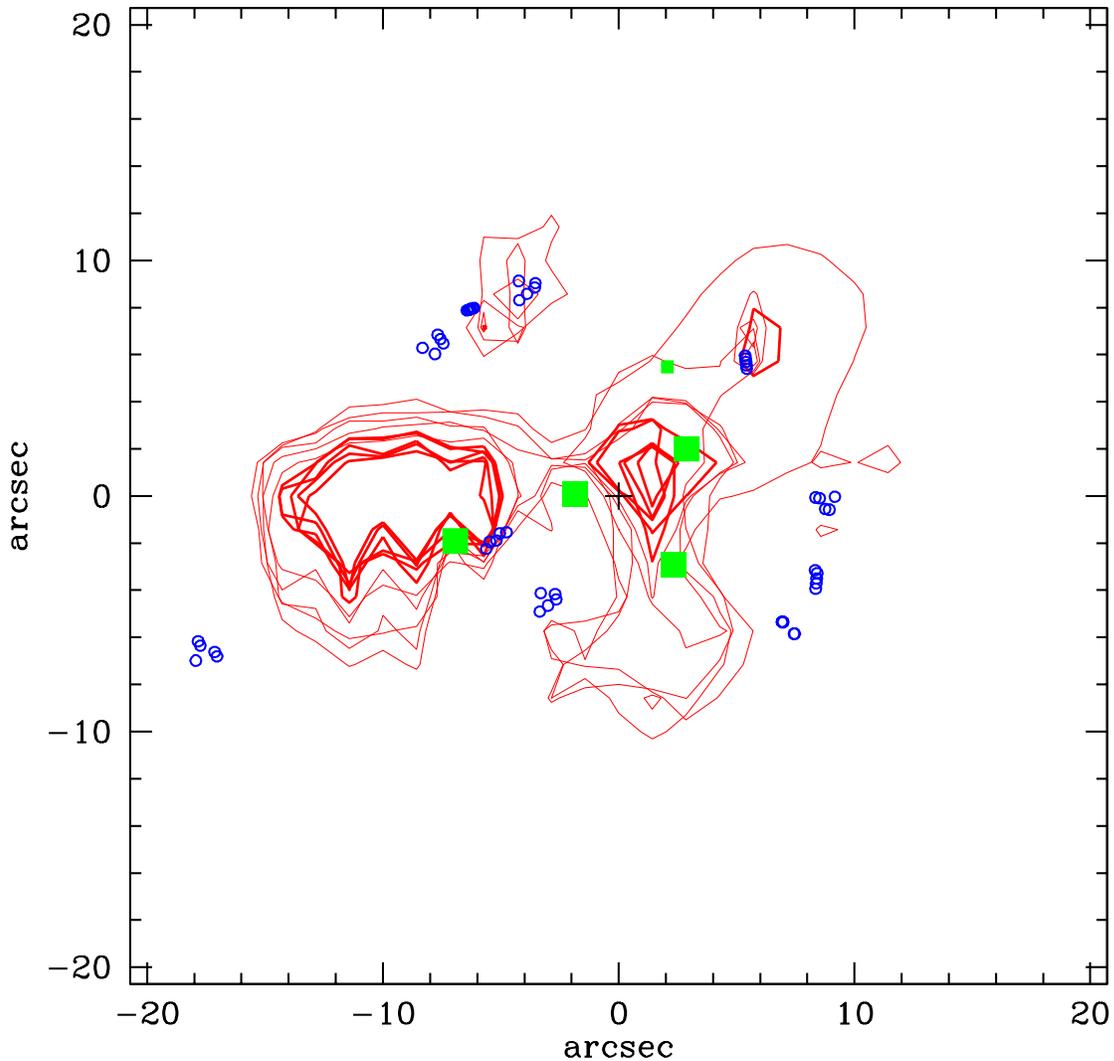}
\vskip-1.5in
\caption{Five overlaid fiducial reconstructions, {\em except} lens image positions have been
randomly scattered within $0''.5$ of the true positions, and are shown as blue empty circles.
Two sets of mass substructure contours per reconstruction are plotted, at 0.5\% (thick lines) and 
2\% (thin lines). The two main features common to all reconstructions are robust against
astrometric uncertainties.} 
\label{fake}
\end{figure}

In addition to the NE-$N1$ clump, the other notable feature of the
reconstructions is that regardless of the input model parameters (see
Figures~\ref{fiducial}-\ref{fake}) there is very little mass excess
associated with $N2$, the centrally located elliptical galaxy. This
suggests that the peak of the mass distribution is not very close to
$N2$.  Here we ask which one of our three trial centres, $C0$, $C1$,
or $C2$ comes closest to the true cluster centre, defined as the
location of the largest mass concentration.  We use the following
procedure.

A PixeLens pixel can assume any mass value provided it agrees with the
image positions and priors, like {\tt dgcone} and {\tt shear}. If
there is no smoothing, the resulting maps usually show a lot of
fluctuations between neighbouring pixels. To avoid that, we impose a
nearest neighbour smoothing constraint by requiring that a pixel
cannot have a mass value larger than twice the average of its eight
neighbours.  The only exception to this rule is the pixel at the
centre of the reconstruction window, which has no imposed upper
bound. It is reasonable to suppose that the closer the reconstruction
centre is to the real centre, the more mass the central pixel will
contain.

From our total (not substructure) mass maps we calculate the fraction
of total mass contained in the central pixel. For the centres $C0$,
$C1$ and $C2$ we calculate that percentage of mass as $1.47\pm 0.25$,~
$2.47\pm 0.22$,~ and $0.79\pm 0.063$ respectively. These values are
averages and rms for the three sets of three maps shown in
Figure~\ref{lots3}. Recall that $C2$ is centered on galaxy $N2$,
whereas $C0$ and $C1$ correspond to 'blank spots' between $N2$ and
$N3$. Since the central pixel of $C2$ contains the least mass, and has
the smallest dispersion of the three cases, it supports our finding in
the earlier Sections that there is very little excess mass associated
with the optical light of elliptical $N2$. Of the three, centre $C1$
is closest to the largest concentration of mass in the cluster.

Finding the centre of the cluster is not the main goal of the paper,
however this Section demonstrates that the uncertainty in the location
of the centre is of order of $1''-2''$, and hence smaller than
fractional uncertainties in the other length measurements entering
eq.~\ref{dispunits}, i.e. $d$ and $s$, therefore when estimating $r$
it is reasonable to use $C0$ as the centre of the cluster, as we have
done in Section~\ref{props}.

\subsection{Adding point masses at the locations of galaxies}\label{pointmass}

To further test the robustness of the NE-$N1$ mass clump we carry out
a reconstruction where PixeLens is allowed to add point masses at
specified pixel locations. In other words, we allow some pixels to
override the nearest neighbour constraint described in
Section~\ref{cent}.

\begin{figure}
%\epsscale{0.95}
%\plotone{smptmass.eps}
%\plottwo{smptmass2v215.pdf}{smptmass1v215.pdf}
\plottwo{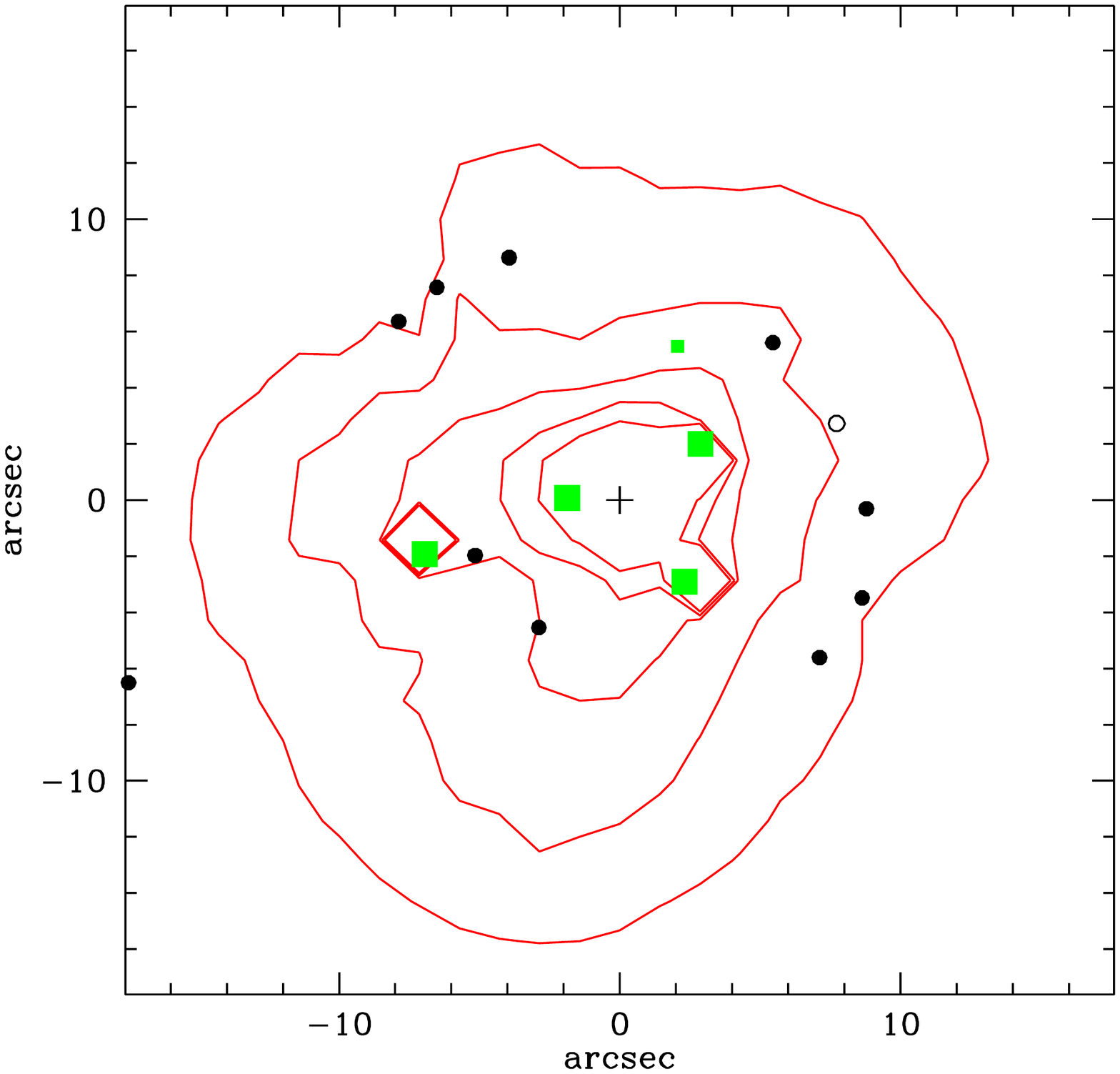}{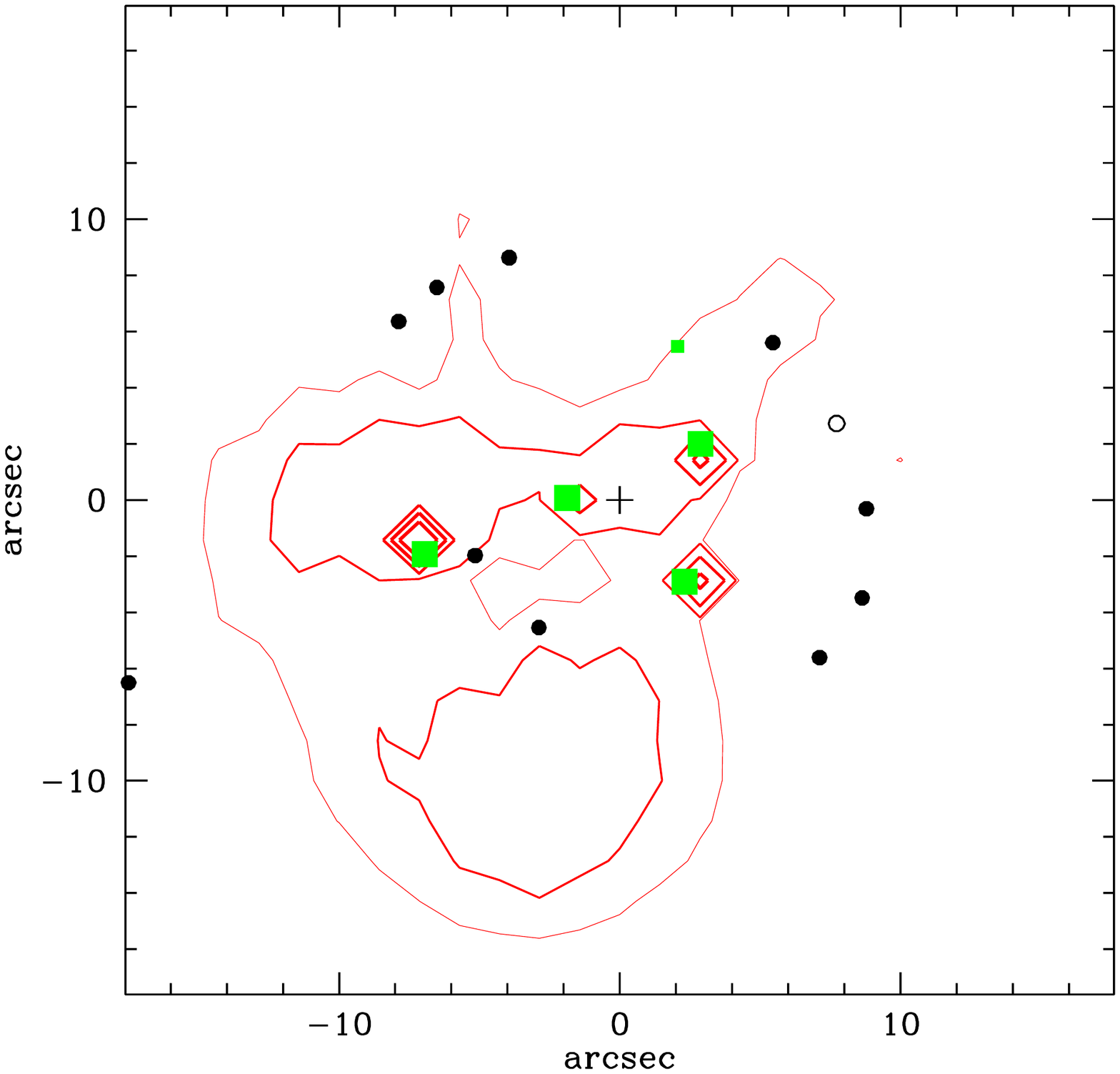}
\vskip-0.5in
\caption{Mass density contours (left) and substructure left after subtracting smooth 
cluster profile (right) of PixeLens fiducial reconstruction, {\it except} here PixeLens 
was specifically allowed to put extra mass at the locations of the four main ellipticals. 
This extra mass is seen as density contours around green squares representing $N1-N4$.
In the left panel the mass countour levels are at 0.9, 1.3, 1.7, 2.1, and 2.5 of critical 
surface mass density for lensing for sources at infinity, while on the right
the contour levels of the mass substructures are at 0.5, 1, 2, 4 and 8\% of the total mass.
Note that the 1\% contour delineates the NE-$N1$ mass clump; see 
Section~\ref{pointmass} for details.} 
\label{ptmass}
\end{figure}

We chose pixels containing the observed galaxies $N1$-$N4$, and use
PixeLens command {\tt ptmass x y area1 area2}, where {\tt x} and {\tt
y} is the location of the point mass, and {\tt area1 area2} are the
lower and upper limits on the area enclosed by its Einstein ring. We
used $0$ and $20$ arcsec$^2$ as representative values for mass
associated with individual galaxies. Figure~\ref{ptmass} shows the
results. The contour levels of the mass substructure are the same as
before, at 0.5, 1, 2, 4, and 8\% of total.

There is a substantial amount of mass associated with individual
galaxies, including $N1$, but a large mass to the upper left of $N1$
is still required, as shown by the 1\% substructure mass contour. In
fact, the NE-$N1$ clump, without the $N1$ point mass, roughly contains
$1-2\;10^{11} M_\odot$, similar to what we find for the fiducial
reconstruction, in Section~\ref{dmx}.  Also consistent with earlier
findings, is that compared to other ellipticals, $N2$ has the least
amount of mass associated with it.

\section{Tests using synthetic mass models}\label{tests}

\subsection{Image $A_{14}$ and the NE-$N1$ mass clump}\label{expl}

In this section we create a series of four synthetic mass
distributions that illustrate why an excess mass lump is needed where
PixeLens places it, to the North East of $N1$.  These synthetic maps
have mass distributions similar enough to the real cluster to make the
comparison meaningful. However the images produced by them do not
attempt to exactly reproduce the observed images.

\begin{figure}
%\epsscale{0.95}
%\plotone{makegal/smfourmassC.eps}
%\plotone{fig9.pdf}
\plotone{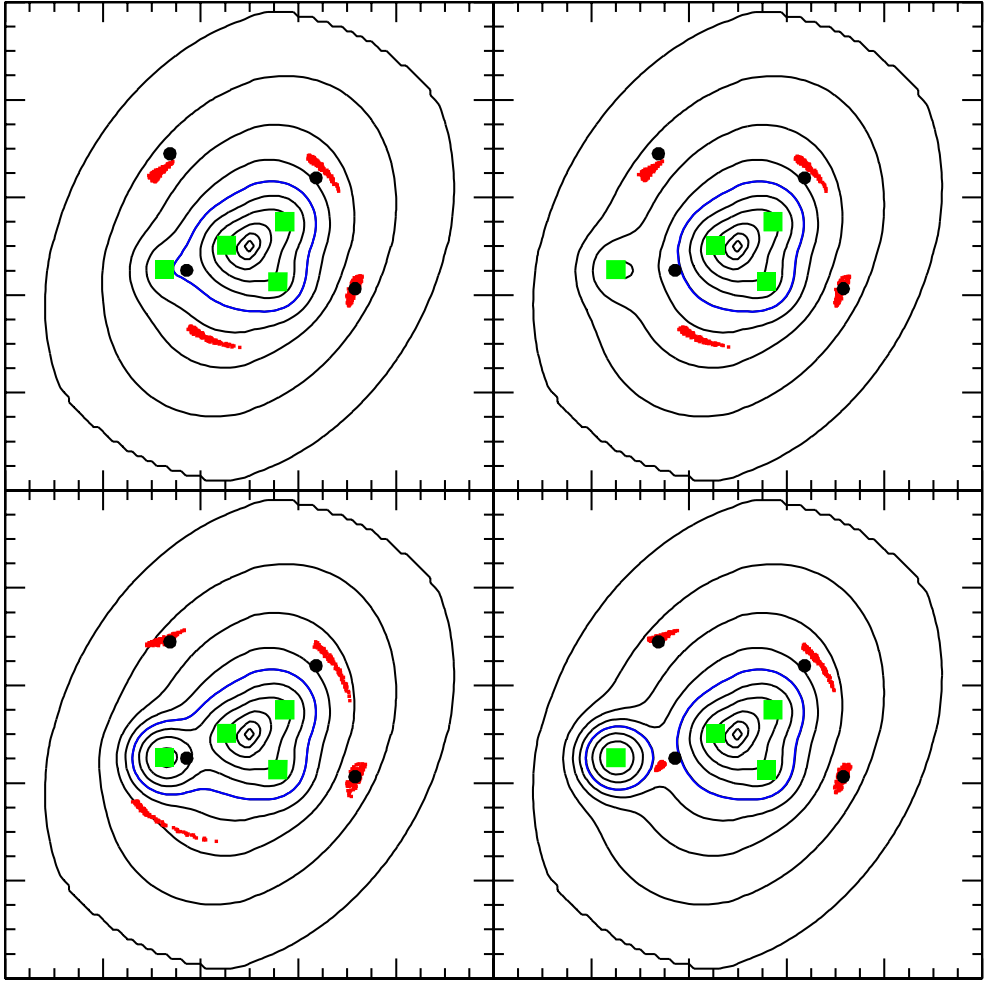}
\vskip-1.5in
\caption{A series of four synthetic mass distributions designed to elucidate why 
free-form PixeLens reconstruction puts a massive secondary clump, the NE-$N1$ clump,
outside of image $A_{14}$. Green squares are the locations of galaxies; their mass contribution
can be judged by how much they distort the density isocontours of the total mass distribution, 
shown as black curves. The blue contour is the critical surface mass density for lensing.
Red islands of points are quad images produced by these mass distributions. Black solid 
points represent the observed images of the $A_{1}$ knot; these are not produced by the 
mass distributions shown in this Figure. Notice that only in the bottom right panel, where $N1$ 
galaxy is made more massive and moved to the upper left, the fourth arriving images of 
the red quads move to the location of the observed $A_{14}$.} 
\label{fourmassC}
\end{figure}

In Figure~\ref{fourmassC} the mass map in the upper left is roughly
what Abell 3827 would look like if mass followed light, in other
words, mass clumps are placed at the observed locations of galaxies
$N1$-$N4$, indicated by the green squares.  The blue contour has the
critical surface mass density for $z_s=0.2$. The black solid dots are
the locations of the observed images of source $A_1$. The four
extended islands of red points mark the locations of images produced
by this synthetic mass configuration.  These images were selected such
that the second arriving images cluster around the observed
$A_{12}$. The other images were unconstrained.

The locations of the 1st, 2nd and 3rd arriving images can be
relatively well reproduced by the mass-follows-light
scenario. However, the 4th arriving images are consistently far from
the observed $A_{14}$. This is the basic reason why mass-follows-light
models do not work.

What does one have to do to (approximately) reproduce $A_{14}$? In the
upper right panel of Figure~\ref{fourmassC} we move the mass clump
associated with $N1$ to the left (East).  The 4th arriving images are
hardly affected. Next, in the lower left panel we keep the mass clump
at its original position but increase its mass by $\times 6$.  The
configuration of images changes but the location of $A_{14}$ is still
not reproduced. Finally, in the lower right panel we move the mass
clump to the left {\em {and}} increase its mass by $\times 6$. Now the
bulk of the 4th arriving images move to where the observed location of
$A_{14}$.  This is basically the mass distribution that PixeLens
finds, with its massive NE-$N1$ clump.

\subsection{Reconstructions of synthetic mass distributions with
PixeLens}\label{synth}

A further test of the veracity of PixeLens's reconstructions is to use
the images created by the synthetic mass distributions of the previous
section as input for PixeLens.

\begin{figure}
%\epsscale{0.99}
%\plottwo{makegal/v192/smpxmassh5.eps}{makegal/v192/smpxmassh4.eps}
%\plottwo{fig10a.pdf}{fig10b.pdf}
\plottwo{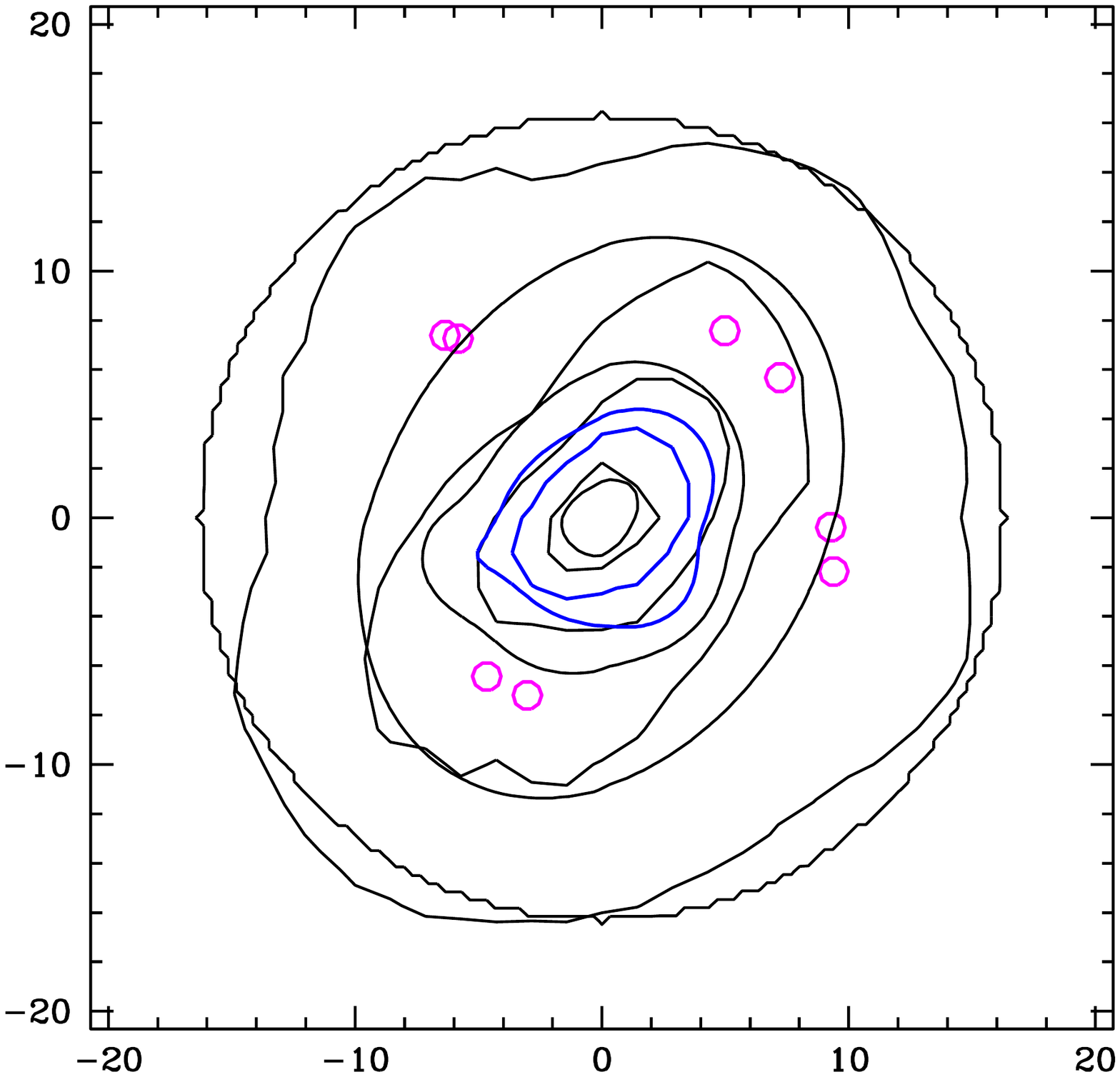}{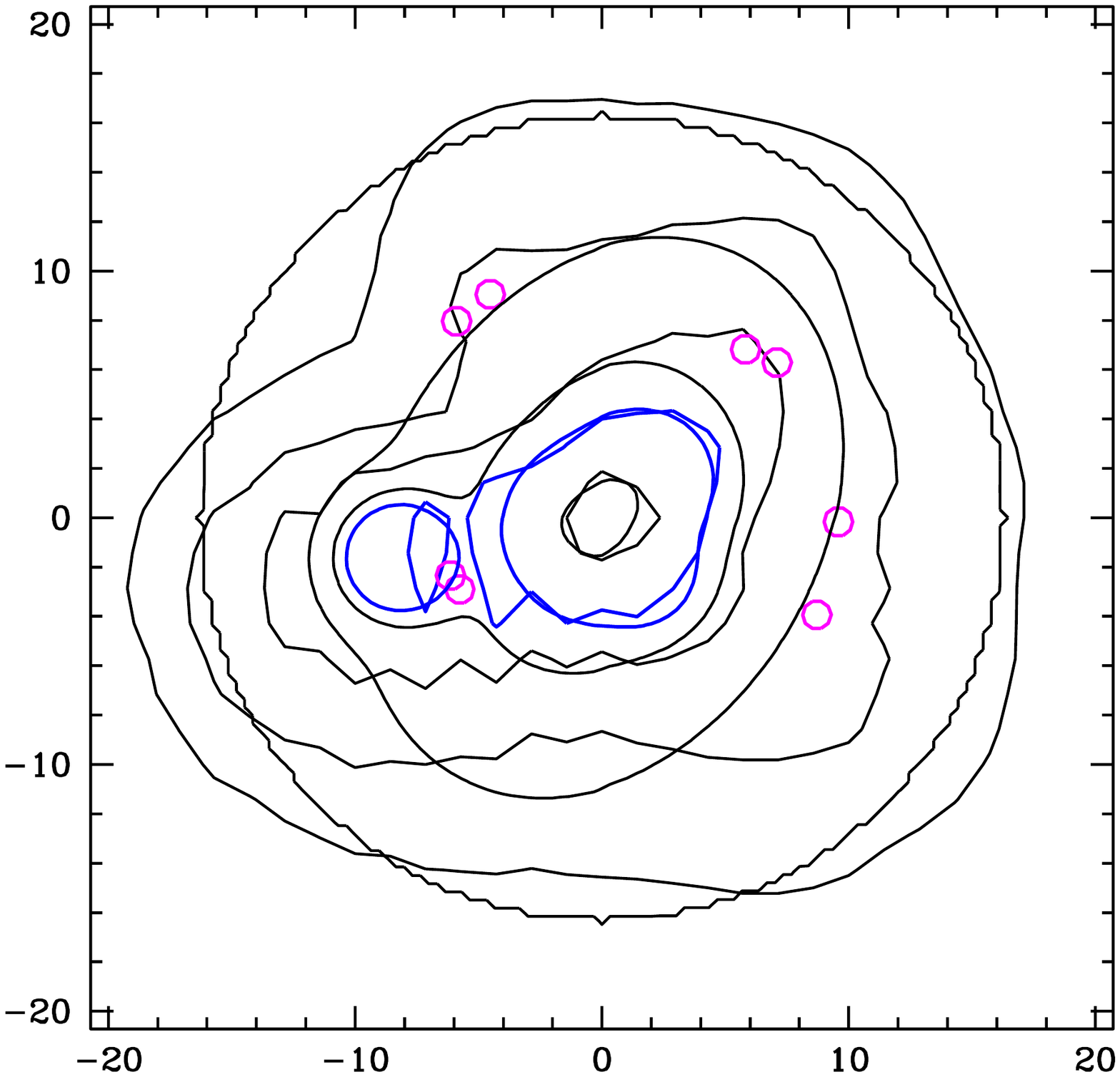}
\vskip-0.5in
\caption{Known synthetic (smooth) and PixeLens reconstructed (jagged) mass contours
of the lens in the upper left and the lower right panels of Figure~\ref{fourmassC}, shown
here in the left and right panels, respectively. The black contours are at 0.25, 0.5, 0.75 
and 2 of critical surface mass density; %for lensing for sources at $z_s=0.2$ 
the blue contour is at 1. The purple empty circles are the images of two quads, 
produced by the synthetic mass distributions and used for the reconstruction.}
\label{pxmass}
\end{figure}

We use the mass distributions of the upper left and lower right panels
of Figure~\ref{fourmassC}. For each of these we take two quads from
the red dots in Figure~\ref{fourmassC}, for a total of eight images to
be used as lensing constraints.  The two panels of Figure~\ref{pxmass}
show the synthetic mass distributions as the smooth contours, while
the jagged contours are the PixeLens reconstructions. The black mass
contours are at 0.25, 0.5, 0.75 and 2 of the critical surface mass
density for $z_s=0.2$, while the blue contour is at the critical
density. The magenta empty circles are the images used in the
reconstruction.

The first of the two synthetic mass models (left panel) is
approximately elliptical and does not have a mass lump to the left of
the centre. PixeLens recovers the mass map quite well; most
importantly it does not create a fictitious massive clump near the 4th
arriving image. The second synthetic mass map (right panel) does have
a mass clump, and PixeLens recovers it well.

\section{Dark matter cross-section}\label{dmx}

The reconstructions of the preceding Sections and the tests with
synthetic lenses show that the two main features of the lensing mass
reconstructions are robust. (1) The NE-$N1$ mass clump is present
consistently in most reconstructions, and it is never centered on
galaxy $N1$. (2) The second, less massive clump is near galaxies $N3$
and $N4$, but avoids $N2$, so that there is no excess mass associated
with that central elliptical.

We interpret (1) to mean that the visible and the dark components of
$N1$ are separated, and we hypothesize that the cause of the
separation is the scenario described in Section~\ref{intro}, namely
that the scattering between the galaxy's and the cluster's dark matter
particles induced the galaxy's halo to lag behind its stellar
component. Other interpretations of the observed separation are
discussed in Section~\ref{other}. Conclusion (2) also speaks to the
dynamically disturbed nature of the cluster core, though there is not
enough information to speculate what happened to the halo of $N2$ and
how the mass clump is related to the nearby $N3$ and $N4$.

We now concentrate on (1), and estimate $\sigma/m$ from
eq.~\ref{dispunits} and the parameters based on the reconstructions
presented in Sections~\ref{effp} and \ref{effs} and summarized in
Section~\ref{props}. We do not consider the rms dispersions quoted in
that Section here, because the main source of uncertainty is not in
the measured properties of the NE-$N1$ clump, but in the dynamics of
the cluster.  For a conservative lower limit on $\sigma/m$ we want to
use the smallest linear distances, $r$, $s$, and $d$ consistent with
the reconstruction, so we use the sky projected quantities, and make
no attempt to estimate the corresponding three-dimensional values.

To obtain $M_{co}$ we measure the projected mass within $r$ from
Figure~\ref{massenc} and halve it to account for the core being a
sphere instead of a long tube; $M_{co}\sim 2.9\; 10^{12} M_\odot$.

One major source of uncertainty is related to the quantity
$M_{co}/r^2$ entering eq.~\ref{dispunits}. First, if $M_{co}/r^2$
depends on radius, i.e. distance from cluster center, then the unknown
shape and ellipticity of $N1$'s orbit and hence how it `samples'
$M_{co}/r^2$ becomes an issue. Second, the radial dependence of
$M_{co}/r^2$ may well be time dependent. From our lensing
reconstruction we can measure how this quantity scaled with radius at
the present epoch. The projected $\Sigma\propto R^{-0.5}$, which
implies that the space density $\rho\propto r^{-1.5}$, and mass
enclosed $M_{co}\propto r^{1.5}$, so $M_{co}/r^2\propto
r^{-0.5}$. This is a relatively weak radial dependence, but it does
not tell us much about the temporal dependence. In future work, these
uncertainties can be constrained through N-body simulations that
sample possible merger histories and have the present day
configuration resembling Abell 3827.

The other large source of uncertainty is also related to cluster
dynamics, and it is $t$, the duration of the encounter, because it
cannot be measured directly from the reconstruction.  A lower
limit on $t$ would be one
dynamical time at the present radius if $N1$, roughly $t\sim 3\times
10^{7}$yr. The upper limit is the age of the Universe,
$t\sim10^{10}$~yrs. The most likely $t$ would probably correspond to
several dynamical times within the extended core region of the
cluster, or $t\sim 10^9$ yr.

Putting all the parameters together in to eq.~\ref{dispunits}
we express the lower limit on the dark matter self-interaction
cross-section as
\begin{equation}
\sigma/m\simgt 4.5\; 10^{-7}
\Bigl(\frac{t}{10^{10} {\rm yr}}\Bigr)^{-2} {\rm cm}^2\,{\rm gm}^{-1}.
\label{estim}
\end{equation}

While this value is safely below the astrophysically measured upper
limit on self-scattering, it is interesting to compare it to the
estimates of the self-annihilation cross-section for dark matter
particles, because in most particle models the latter are expected to
be larger than the scattering cross-sections.  To facilitate the
comparison we express our result in terms of cross-section per
particle of mass $m$, and attach a subscript to indicate that this
refers to self-scattering,
\begin{equation}
\sigma_{scat}\simgt
8.1\times 10^{-29} \Bigl(\frac{t}{10^{10}{\rm yr}}\Bigr)^{-2}
\Bigl(\frac{m}{100\;{\rm GeV}}\Bigr){\rm cm}^2.  \label{estim2}
\end{equation}
If the dark matter particle is a thermal relic, its self-annihilation
cross-section can be expressed in terms of dark matter density and
typical velocity \citep{bert05},
\begin{equation}
\sigma_{ann,th}\approx 10^{-33} \Bigr(\frac{\langle
v\rangle}{300\;{\rm{km}\;{\rm s}^{-1}}}\Bigr)^{-1}
\Bigl(\frac{\Omega_{dm}h^2}{0.1}\Bigr)^{-1}{\rm cm}^2.
\end{equation}
A different constraint comes from considering the observed abundance
of annihilation by-products. For neutrinos, \cite{beac07} get an upper
bound which is more stringent than provided by other experimental and
observational considerations,
\begin{equation}
\sigma_{ann,up}\approx
3\times 10^{-29} \Bigr(\frac{\langle v\rangle}{300\;{\rm{km}\;{\rm
s}^{-1}}}\Bigr)^{-1}{\rm cm}^2
\end{equation}
for a range of particle masses between $1$ and $10^4$ GeV. While our
result is well above (i.e. inconsistent with) the thermal relic
estimate, $\sigma_{ann,th}$, for particle masses above $1$~MeV, it is
comparable to $\sigma_{ann,up}$, especially if particle mass is
somewhat below $100$ GeV.

\section{Other interpretations}\label{other}

Here we offer other interpretations for the separation between the
NE-$N1$ mass clump and the elliptical galaxy $N1$.

(i) The NE-$N1$ mass clump could belong to the line of sight
structures, and not the cluster core, or galaxy $N1$. In that case the
alignment of NE-$N1$ clump with the cluster ellipticals is not
expected. However, there are no additional light concentrations in the
direction of the NE-$N1$ clump, so if this mass is external to the
cluster core, it would still need to have a very high mass-to-light
ratio.

(ii) The NE-$N1$ mass clump could be primarily due to the baryonic gas
associated with galaxy $N1$, which has been separated from its parent
galaxy by ram pressure of the cluster gas. However, in the cluster's
core one would expect the gas of individual ellipticals to have been
stripped and dispersed on short time-scales, and in the case of a late
stage merger, as in Abell 3827, long before the galaxies of the
merging clusters reach the new common cluster core.  Observations of
other massive clusters do not show X-ray emitting gas associated with
individual galaxies in the core.  Examination of an archival Chandra
image of Abell 3827 also shows no discernible substructure 
(Proposal ID 08800836). %PI Zappacosta

(iii) Because the visible and dark components of $N1$ have different
spatial extents, the gravitational field gradient of the cluster could
have resulted in different forces acting on the two components, which
could lead to the observed separation. However, it is hard to imagine
how tidal forces would lead to such a large offset. Another
possibility is that the stellar component of $N1$ is wobbling around
the centre of its dark matter's potential well. Separation due to this
effect have been observed, but only for the central cluster
galaxies. $N1$ is not the central elliptical; $N1$ and NE-$N1$ clump
are both about $20$~kpc from cluster's largest mass concentration
(Section~\ref{cent}).

(iv) Finally, because the lensed images are faint, extended and close
to bright ellipticals, the identification and location of images could
be partially wrong. Since image positions are the most important input
for PixeLens mass reconstruction, misidentified images could lead to
wrong recovered mass maps. Additional observations would help; for now
we note that random astrometric uncertainties, of order $0''.5$, in
the image positions do not affect the main features of the
reconstruction (Section~\ref{uncert}).

\section{Conclusions}

In this paper we argue that if the dark matter self-scattering
cross-section is only a few orders of magnitude below its most
stringent astrophysically determined upper limit, $\sigma/m\sim
0.02$~cm$^2$~g$^{-1}$, then it should be detectable in the cores of
massive galaxy clusters. The dark matter particles of individual
galaxies that have been orbiting the cluster centre for about a Hubble
time will be scattering off the dark matter of the cluster, resulting
in a drag force, not experienced by the visible stellar component of
the galaxies. Over time this drag will amount to a spatial separation
between the dark and visible components of the galaxy, whose magnitude
is estimated by eq.~\ref{dispunits}.

We suggest that at least one elliptical in the core of Abell 3827 is
an example of such a scenario. Our free-form lensing reconstructions
show that there a massive dark clump, the NE-$N1$ clump, located about
$3''$, or $6$ kpc from the centre of light of $N1$.  We performed many
mass reconstructions to test the robustness of the NE-$N1$ clump, and
quantified the uncertainty in its properties. A series of synthetic
mass models, presented in Section~\ref{tests}, further support the
existence of the clump and its separation from the visible component
of $N1$.

We interpret this clump as the dark matter halo of $N1$ which is
lagging behind its stellar counterpart. If this interpretation is
correct, we estimate the lower limit on the cross-section as given in
eq.~\ref{estim}. The largest source of systematic uncertainty in this
equation relates to the unknown dynamics of the cluster and central
galaxies, the shape and inclination of the orbit of $N1$, and the true
three dimensional distances and separations within the core.

In Section~\ref{other} we note that other interpretations of the
observed separation cannot be ruled out at this point. Future
observations as well as numerical simulations should be able to
differentiate between the various scenarios, and also quantify the
uncertainties associated with the dynamical state of the cluster.
%For example, scenario (iii) of Section~\ref{other} can be %tested
using purely collisionless simulations. The stellar and dark matter
components of %a galaxy can be represented by two, initially
superimposed, mass distributions of different %scale-lengths. If after
several orbits around the cluster centre the two components 'wobble'
%differently and their centres are displaced from each other, then
there is no need for %$\sigma/m$ to be non-zero. The scenario with
non-zero $\sigma/m$, Section~\ref{intro}, %can also be simulated to
check if an observable separation can develop over a Hubble time.

As already mentioned, Abell 3827 is special in having very favorable
parameters for the detection of the dark matter-light offset. First,
A3827 is a very massive cluster, hosting several massive ellipticals
in the very core. \cite{car10} note that within a similar radius,
Abell 3827 is slightly more massive than Abell 1689 \citep{lim07}, a
cluster with over 100 strong lensing features. They also note that the
central galaxy of Abell 3827 is perhaps the most massive cD galaxy in
the local Universe.  Second, the fortuitous location of the background
source A produces strong lensing features that are optimal in
configuration and distance for mass reconstruction in the most
relevant regions; the low redshift of Abell 3827 and source A ensure
that the images are close to the cluster centre, $\simlt 10''$. These
factors make it unlikely that there are many other clusters out there
where the offset would be detectable.

\acknowledgements

LLRW would like to thank Henry Lee for discussing with her the images
of this unusual object prior to publication, Keith Olive for
discussions about particle dark matter, and Larry Rudnick for help
with Chandra data archive.

{}

\end{document}